\documentclass[12pt,letterpaper]{article}

\usepackage{lmodern}
\usepackage{latexsym,multirow}
\usepackage{amssymb,amsmath,bm}
\usepackage{bigints}
\usepackage{array}
\usepackage{longtable}
\usepackage{booktabs}       
\usepackage{bbm} 
\usepackage{graphicx,float}
\usepackage[color,all,import,arrow]{xy}
\usepackage{enumerate}
\usepackage{verbatim}
\usepackage{lscape}
\usepackage{mathtools}
\usepackage{fancyvrb}
\usepackage{pdfpages}

\usepackage[format=plain,
            labelfont={bf,it},
            textfont=it]{caption}

\usepackage{longtable}

\usepackage{placeins}

\usepackage[natbib=true,minnames=1,maxnames=2,backend=biber,style=authoryear-luh-ipw]{biblatex} 
\usepackage{atbegshi}
\AtBeginDocument{\AtBeginShipoutNext{\AtBeginShipoutDiscard}}

\usepackage{dcolumn}
\newcolumntype{.}{D{.}{.}{-1}}
\newcolumntype{d}[1]{D{.}{.}{#1}}

\graphicspath{{./}}

\usepackage{subcaption}
\DeclareCaptionLabelFormat{step}{Step \textbf{#2)}} 

\usepackage[T1]{fontenc}
\usepackage{tikz}
\usepackage{qtree}
\usepackage{comment}
\usetikzlibrary{arrows}
\usetikzlibrary{positioning}
\usetikzlibrary{matrix}
\usetikzlibrary{bayesnet}
\usetikzlibrary{decorations.pathreplacing,calligraphy} 

\usepackage[ruled,vlined]{algorithm2e}

\usepackage{theorem}
\theoremstyle{plain}
\theoremheaderfont{\scshape}

\usepackage{chngpage}

\usepackage{rotating}

\usepackage{arydshln}

\usepackage[compact]{titlesec}

\usepackage{setspace}
\usepackage[bottom]{footmisc}
\allowdisplaybreaks

\newcommand\spacingset[1]{\renewcommand{\baselinestretch}%
{#1}\small\normalsize}

\newcommand{\blind}{0}

\newcommand{\bX}{\mathbf{X}}

\newcommand{\bY}{\mathbf{Y}}

\usepackage{xr}
\usepackage[pdftex, bookmarksopen=true, bookmarksnumbered=true,
pdfstartview=FitH, breaklinks=true,
urlbordercolor={0 1 0}, citebordercolor={0 0 1}]{hyperref}
\usepackage{colortbl}

\usepackage{CJKutf8}

\begin{document} 

\newcommand{\tit}{
Linking Datasets on Organizations Using Half a Billion Open-Collaborated Records
}

%
\spacingset{1.25}

\if0\blind

{\title{\bf\tit\thanks{
To appear in {\it Political Science Methods and Research (PSRM)}. We thank Beniamino Green, Kosuke Imai, Gary King, Xiang Zhou, members of the Imai Research Workshop, and two anonymous reviewers for valuable feedback. We would also like to thank Neil Arora, Danny Guo, Gil Tamir, and Xiaolong Yang for excellent research assistance. We also thank Daniel Carpenter for making this project possible. 
}

  \author{    
      Brian Libgober
\thanks{Assistant Professor of Political Science and Law, Department of Political Science, Northwestern University. Email:
\href{mailto:jbrian.libgober@northwestern.edu}{brian.libgober@northwestern.edu}; URL:
\href{https://brianlibgober.com/}{BrianLibgober.com};  ORCID: 0000-0001-9638-0228.}
\and 
Connor T. Jerzak\thanks{Assistant Professor, Department of Government, The University of Texas at Austin.  Email: \href{mailto:connor.jerzak@austin.utexas.edu}{connor.jerzak@austin.utexas.edu}; URL:
\href{https://connorjerzak.com}{ConnorJerzak.com}; ORCID: 0000-0003-1914-8905.
}
\date{
    \today
}
}
}

\fi

\if1\blind
\title{\bf \tit}
\fi

\maketitle

\pdfbookmark[1]{Title Page}{Title Page}

\thispagestyle{empty}
\setcounter{page}{0}
         
\begin{abstract}
\noindent Scholars studying organizations often work with multiple datasets lacking shared identifiers or covariates. In such situations, researchers usually use approximate string (``fuzzy'') matching methods to combine datasets. String matching, although useful, faces fundamental challenges. Even where two strings appear similar to humans, fuzzy matching often struggles because it fails to adapt to the informativeness of the character combinations. In response, a number of machine learning methods have been developed to refine string matching. Yet, the effectiveness of these methods is limited by the size and diversity of training data. This paper introduces data from a prominent employment networking site (LinkedIn) as a massive training corpus to address these limitations. By leveraging information from the LinkedIn corpus regarding organizational name-to-name links, we incorporate trillions of name pair examples into various methods to enhance existing matching benchmarks and performance by explicitly maximizing match probabilities. We also show how relationships between organization names can be modeled using a network representation of the LinkedIn data. In illustrative merging tasks involving lobbying firms, we document improvements when using the LinkedIn corpus in matching calibration and make all data and methods open source.

\vspace{.25cm} 

\end{abstract}


\clearpage
\spacingset{1}

\section*{Introduction}
As large datasets on individual political behavior have become more common, scholars have focused increasing attention on the methodological problem of linking records from different sources \citep{ruggles2018historical,Enamorado2019,Herzog2010,Larsen2001}. Record linkage is an all too common task for researchers building datasets. When a unique identifier, such as a social security number, is shared between data collections and made available to researchers, the problem of record linkage is significantly reduced. Errors in linkage, presumably rare, may be regarded as sources of noise. In cases where unique identifiers like social security numbers are not available, recent literature has developed probabilistic linkage algorithms that can find the same individual in two datasets using stable characteristics such as birth year and race, or even mutable characteristics such as address \citep{Enamorado2019}. The rise of such techniques has paved the way for research that would have been costly or impossible to conduct in previous eras \citep[e.g.,][]{Figlio2014a,Bolsen2014,Hill2017}.

These new techniques have had less of an impact so far on scholarship concerning organizational entities, such as corporations, universities, trade associations, think tanks, religious groups, nonprofits, and international associations---entities that are important players in theories of political economy, American politics, and other (sub-)fields. Like researchers on individuals, scholars studying organizations also seek to combine multiple data streams to develop evidence-based models. However, in addition to lacking shared unique identifiers, datasets on organizations {\it also} often lack common covariate data that form the basis for probabilistic linkage algorithms. Therefore, scholars must (and do) rely heavily on exact or fuzzy string matching based on names to link records on organizations---or, alternatively, bear the significant costs of manually linking datasets.

To take an example from the applied political science literature, \citet{Crosson2020} compare the ideology scores of organizations with political action committees (PACs) to those without. Scores are calculated from a dataset of position-taking interest groups compiled by a nonprofit (Maplight). The list of organizations with PACs comes from Federal Election Commission (FEC) records. Maplight and the FEC do not refer to organizations using the same names. There is no covariate data to help with linkage. The authors state that matching records in this situation is ``challenging'' (p.~32) and consider both exact and fuzzy matching as possibilities. Ultimately, they perform exact matching on names after considerable preprocessing because of concerns about false positives, acknowledging that they do not link all records as a result. Indeed, the authors supplement the 545 algorithmic matches with 243 additional hand matches, implying that the first algorithmic approach missed about one in three correct matches.

The challenge faced by \citet{Crosson2020} is typical for scholars studying organizations in the US or other contexts. Given the manageable size of their matching problem, the authors are able to directly match the data themselves and bring to bear their subject matter expertise. In many cases, where the number of matches sought is not in the hundreds but in the thousands, practical necessity requires using computational algorithms like fuzzy matching or hiring one or more coders (e.g., undergraduates or participants in online markets such as Amazon's Mechanical Turk). 

Both string matching and reliance on human coders have limitations. Even though string distance metrics can link records whose identifiers contain minor differences, they do not optimize a matching quality score and have trouble handling the diversity of monikers an organization may have. For example, ``JPM'' and ``Chase Bank'' refer to the same organization, yet these strings share no characters. Likewise, string matching and research assistants would both have difficulty detecting a relationship between Fannie Mae and the Federal National Mortgage Association. Such complex matches can be especially difficult for human coders from outside a study's geographic context, as these coders may lack the required contextual information for performing such matches.

Methodologists have started tackling the challenges that researchers face in matching organizational records.  \citet{kaufman_klevs_2021}, for example, propose an adaptive learning algorithm that does many different kinds of fuzzy matching and uses a human-in-the-loop to adapt possible fuzzy-matched data to the researcher's particular task. While this approach represents an improvement over contemporary research practices, an adaptive system based on fuzzy matching still requires researchers to invest time in producing manual matches and may also struggle to make connections in the relatively common situation where shared characters are few and far between (e.g.,~Chase Bank and JPM) or where characters are shared but the strings have very different lengths (e.g., Fannie Mae and Federal National Mortgage Association). Scholars are also turning to large language models for performing name linkage tasks \citep{agrawal2022large}; however, these large language models have not been fine-tuned on match tasks, so they may struggle to produce matches similar to how fuzzy matching does. 

In this paper, we leverage a data source containing half a billion open-collaborated records from the employment networking site LinkedIn, which can serve as a resource for scholars who seek to link records about organizations. We show how this dataset can assist in three distinct kinds of record linkage methods---the first approach based on machine learning, the second based on network analysis and community detection, and the third based on a combination of network and machine learning methods. Intuitively, each approach uses the combined wisdom of millions of human beings with first-hand knowledge of these organizations. Our argument is that this combined wisdom from trillions of real-world name-pair examples can, if incorporated into a given linkage strategy, improve matching performance at relatively little cost. 

In what follows, Section \ref{s:Description} describes the massive training dataset we constructed from a scrape of LinkedIn. Section \ref{s:ModelingHighLevel} describes how the LinkedIn data can be used to improve linkage given two distinct representations of the data stream. Section \ref{s:Illustration} illustrates the use of these linkage methods on three tasks revolving around the role of money in politics. Section \ref{s:Discussion} and Section \ref{s:Conclusion} conclude. An open-source package (\Verb|LinkOrgs|) implements the methods we discuss. We make the massive LinkedIn name-match corpus available in a Dataverse (\href{doi.org/10.7910/DVN/EHRQQL}{\url{doi.org/10.7910/DVN/EHRQQL}}).

\section{Employment Networking Data as a Resource for Scholars of Organizational Politics}\label{s:Description}

In this section, we explain how records created by users on LinkedIn, a leading professional networking platform, hold a wealth of information relevant for researchers studying organizational politics, particularly in the ubiquitous yet challenging task of assembling datasets.

The key insight for the data asset we built is that LinkedIn users provide substantial information about their current and previous employers. For the sake of our illustration, we will use a near census of the publicly visible LinkedIn network circa 2017, which we acquired from the vendor \Verb|Datahut.co|. Researchers do have the legal right to scrape this website and use the updated corpus (as the Ninth Circuit Court of Appeals established in \textit{HIQ Labs, Inc., v. LinkedIn Corporation} (2017)). That said, these data do not come cheaply, and, informally, it seems to us that costs have increased as a result of greater investment in anti-scraping technology by site owners in the wake of the decision. Although we do not have a more recent scrape available to us at this time, there are vendors with more recent versions (e.g., using LinkDB \citep{Goh2022}). We expect over time that the approaches we take to the 2017 data will be applicable to later scrapes as they become available to the field. The dataset we use contains about 350 million unique public profiles drawn from over 200 countries---a similar size and coverage to LinkedIn's estimates reported during its 2016 acquisition by Microsoft.\footnote{At the time of acquisition, 433 million total members and 105 million unique visitors per month were reported \citep{Microsoft2016AcquireLinkedIn}. We are not able to find authoritative counts of the number of publicly visible profiles.}

To construct a linkage directory for assisting dataset merges, we here use the professional experience category posted by users. In each profile on LinkedIn, a user may list the name of their employer as a free-response text. We will refer to the free-response name (or ``alias'') associated with unit \(i\) as \(A_i\). In this professional experience category, users also often post the URL link to their employer's LinkedIn page, which we can denote as $U_i$. This URL link serves as an identifier for each organization.

\begin{table}[ht]
\caption{Illustration of source data using three public figures.\label{tab:excerpt}}
\resizebox{\textwidth}{!}{%
\begin{tabular}{l  l  l  l  l  l}
\hline \hline 
\textit{Name} & \textit{Title} & \textit{Organization} & \textit{Organization URL Path} (\Verb|linkedin.com/company/|) & \textit{Start date} & \textit{End date} \\
\hline
Michael Cohen & EVP \& Special Counsel to Donald J. Trump & The Trump Organization & \url{the-trump-organization} & 20070501 & 20170418 \\
Allen Weisselberg & EVP/CFO & The Trump Organization & \url{the-trump-organization} &  & 20170316 \\
Michael Avenatti & Founding Partner & Eagan Avenatti, LLP &  & 20070101 & 20170318 \\
Michael Avenatti & Chairman & Tully's Coffee & \url{tully's-coffee} & 20120101 & 20170318 \\
Michael Avenatti & Attorney & Greene Broillet \& Wheeler, LLP &  & 20030101 & 20070101 \\
Michael Avenatti & Attorney & O'Melveny \& Myers LLP & \url{o'melveny-\&-myers-llp} & 20000101 & 20030101 \\
\hline \hline 
\end{tabular}
}
\end{table}

Table \ref{tab:descriptive} provides descriptive statistics about the scope of the dataset as it relates to organizational name usage. The statistics reveal that, on average, users refer to organizations in about three different ways and that, on average, each of the 15 million aliases links to slightly more than one organizational URL. The table also notes that there are more than $10^{14}$ alias pairs. The database contains a large number of ways names can refer to the same or different organizational URLs.

\begin{table}[ht]
\centering
\caption{Descriptive statistics for the LinkedIn data.}\label{tab:descriptive}
\begin{tabular}{lc} 
\hline \hline 
\textit{Statistic} & \textit{Value} \\
\hline 
{\# unique aliases} & 15,270,027 \\
{\# unique URLs} & 5,950,995 \\
{Mean \# of unique aliases per URL} & 2.88 \\
{Mean \# of URL links per unique alias} & 1.12 \\
{Total \# of alias pair examples} & $>10^{14}$ \\
\hline \hline 
\end{tabular}
\end{table}

\section{Individual and Ensemble Approaches to Record Linkage Using the LinkedIn Corpus}\label{s:ModelingHighLevel}

We begin our discussion of how to best use the LinkedIn corpus with an example. Suppose we have two datasets, $\bX$ and $\bY$. $\bX$ contains data about Wells Fargo Bank, JP Morgan Chase Bank, and Goldman Sachs. $\bY$ contains data about Wells Fargo Advisors, Washington Mutual (at one time a wholly owned subsidiary of JP Morgan Chase), and Saks Fifth Avenue. Ideally, manual linkage would successfully match Wells Fargo Bank with Wells Fargo Advisors and perhaps even JP Morgan Chase Bank with Washington Mutual, while rejecting all other matches---including between Goldman Sachs and Saks Fifth Avenue, despite some passing phonetic similarity between the names.  

\begin{figure}
\centering
\includegraphics[width=0.7\linewidth]{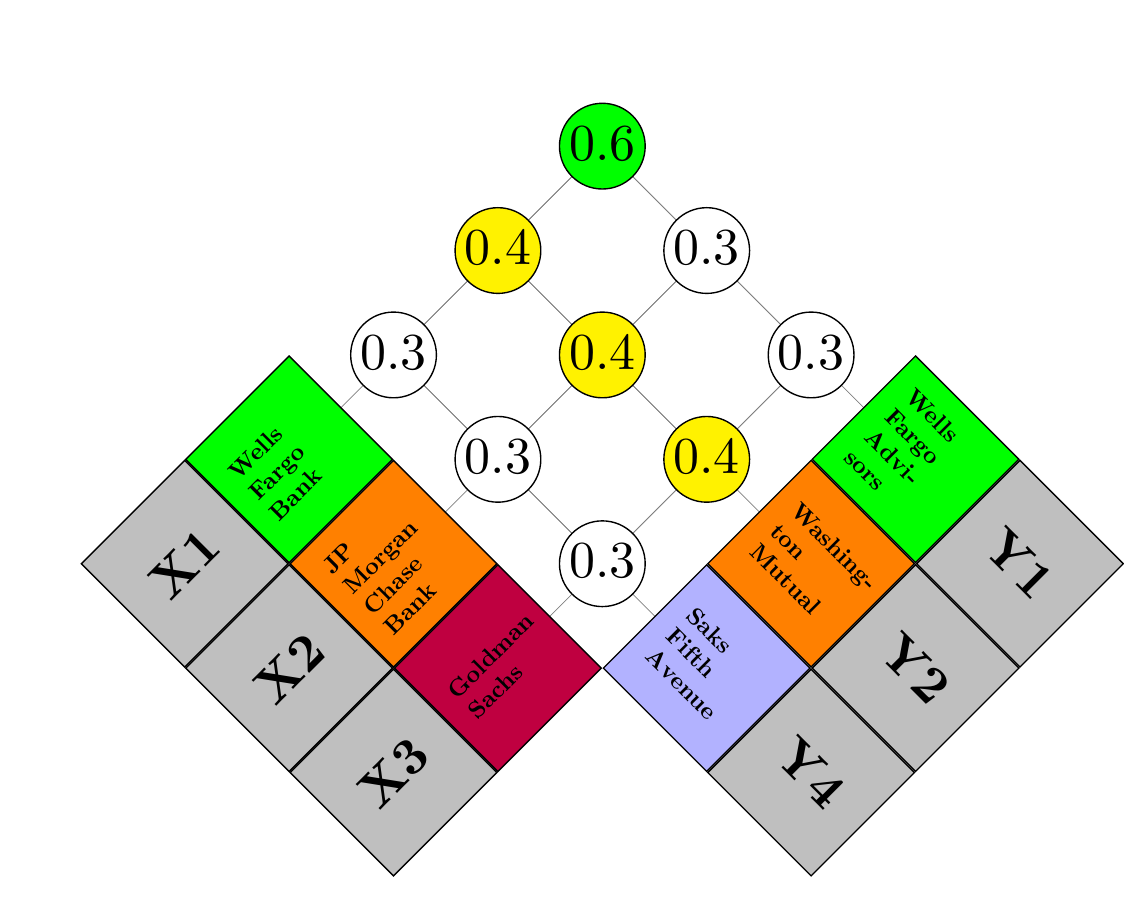}
\caption{Checkered flag diagram describing the organizational linkage problem. \label{fig:basic-problem}}
\end{figure}

Figure \ref{fig:basic-problem} presents a checkered flag diagram illustrating our task. Names in the $\bX$ dataset are on the left side. Names in the $\bY$ dataset are on the right. Each pair of names is represented by a node. To perform linkage, investigators apply an algorithm that assigns scores to all the nodes. They then consider a node to represent a match if it clears some numeric threshold (or, alternatively, investigators can re-weight links in accordance with their match probability). There are many possible functions to score pairs. For example, exact matching scores each node as 1 if $A_i$ and $A_j$ are equal, 0 if unequal, accepting all pairs where the score is 1. In this example, exact matching would fail to link any of the organizations. The figure presents scores using the fuzzy matching approach, in this case, with the Jaccard metric.

These scores present a familiar trade-off. If a cutoff of $0.7$ is adopted, nothing matches. If $0.5$ is selected, then Wells Fargo Bank successfully matches to Wells Fargo Advisors, but JP Morgan Chase Bank does not match to Washington Mutual. If a cutoff of $0.35$ is selected, all the correct matches are included but also several wrong ones. If a cutoff of $0.2$ is selected, everything matches everything. There are no perfect options.

The familiar trade-off facing researchers in this example comes about in part because the scores are too similar between pairs that we do and do not wish to match. Our focus is on algorithmic interventions that can produce scores that make it easier to distinguish between pairs that should match and those that should not at any particular cutoff. 

While ultimately we will propose an ensemble of two distinct approaches, we begin here by discussing the intuition underlying each. The first idea is that, to the extent that an algorithm trained on the LinkedIn corpus can reward similarities in latent meanings (e.g., ``bank'' and ``mutual'') and punish dissimilarities (e.g., ``bank'' and ``avenue''), it stands a good chance of improving upon existing character similarity methods. Machine learning approaches to this task are appealing, and we focus on applying these methods using the LinkedIn network. Despite the increasing sophistication of learning algorithms, these methods do have limitations, as the words in a name do not always have semantic value. Indeed, without specialized domain knowledge, it may be easy to miss matches between organizations having multiple acronyms. To account for this limitation, an approach that utilizes community detection algorithms is desirable and complements the first approach by leveraging alias-to-URL links more explicitly. We provide more details on each approach below and conclude by describing how to unify them in an ensemble.

\subsection{Machine Learning Approach}\label{ss:ml}
Machine learning continues to make so many advances that it is becoming hard to choose, let alone justify as best, any particular framework. The future will no doubt yield improvements in any machine learning approach for modeling name match probabilities, as the rapid progress in large language models has made apparent \citep{wei2022emergent,jiang2023mistral}. That said, to make progress we need to make and explain our choices.

We set up the machine learning problem as the task of learning a function, $f_n$, that maps a textual alias, $a$, to a point in a high-dimensional, real-valued vector space. The distance between two aliases, $a$ and $a'$, is to be calculated as $||f_n(a) - f_n(a')||$. We are mindful that one major benefit of learning a map from the space of strings to numerical vectors is faster matching. String similarity algorithms, such as the Jaccard algorithm, typically require an operation on each combination of entries that one wishes to match in sets $\bX$ and $\bY$; the calculation of a single score can be quite time-consuming. By contrast, applying $f_n$ to all the entries of $\bX$ and $\bY$ generates two sets of points in a vector space. Calculating a distance between all pairs of points is typically a much faster computation than the equivalent string distance calculation on all the pairs. 

The function, $f_n$, that our algorithm will learn is, to a degree, a black box. It optimizes over many parameters by seeking to best fit some target outcome; hence, the way we structure the target influences the ultimate algorithm we produce. Perhaps the simplest approach to setting up an outcome would be to look at the set of all alias pairs, $\mathcal{P} = \{\mathcal{A}_i\} \times \{\mathcal{A}_j\}$, and assert that two aliases are linked if two people used those aliases to refer to the same URL and not linked if no one ever did. A limitation of this ``lookup table''-like approach is that it has no sensitivity to the number of links in the data, so a person who mistakenly writes ``Goldman Sachs'' as their employer and links to the Saks Fifth page would get equal weight to the much more common case where employees write that they work at ``Goldman Sachs'' and link to the Goldman page.

Incorporating information about the relative number of links is clearly desirable, but requires care. As a starting point, we borrow ideas from naive Bayes classifiers to calculate a probability that two aliases are indeed true matches using the depth of ties between links. In Section \ref{ss:DeriveOutcomeML}, we explain assumptions that would allow the interpretation of this outcome variable as a probability, written as: 
\begin{equation}
    \Upsilon_{ij}=\Pr\left(\textrm{$i$ and $j$ match} | A_i=a,A_j=a'\right) = \sum_{u\in\mathcal{U}} \Pr(U_i = u \mid A_i=a)\Pr(U_j = u \mid A_j=a') \label{eq:bayes}
\end{equation}
\noindent To explicate this formula, note that for each URL $u$, the term $\Pr(U_i = u \mid A_i=a)\Pr(U_j = u \mid A_j=a')$  reflects the proportion of occurrences of $u$ given alias $a$ times the proportion of occurrences of $u$ given the alias $a'$. As an illustration, consider a profile URL like \url{LinkediIn.com/company/wellsfargo} and the two aliases, ``JP Morgan Chase Bank'' and ``Wells Fargo Advisors.'' Whenever ``JP Morgan Chase Bank'' is used, it generally occurs with a different company profile, so this particular URL contributes little to the overall probability even though ``Wells Fargo Advisors'' almost always links to this particular profile URL. By contrast, if ``Wells Fargo Bank'' and ``Wells Fargo Advisors'' both typically link to this same profile page, then the probability of a match will be calculated as high. The overall loss function we seek to minimize is
\begin{align}
\textrm{\it Loss} &= \sum_{i}\sum_{j}\;\textrm{KL}( \widehat{\Upsilon}_{ij},\; \Upsilon_{ij}),
\end{align}
where the KL divergence computes the distance in probability space between $\widehat{\Upsilon}_{ij}$, the predicted match probability, and $\Upsilon_{ij}$, the match probability as computed using the LinkedIn corpus. 

We now discuss how we structure the $f_n$ function that ultimately generates $\widehat{\Upsilon}_{ij}$ (for details, see Section \ref{ss:ModelingDetails}). Our approach builds from work on the vector representations of words \citep{mikolov2013distributed}. In our case, we build a model for organizational name matches from the characters on up.\footnote{This sequential approach pays more attention to the order of characters/words than the traditional bag-of-words/characters approaches that historically saw wide use in political science text analysis; for discussion of word-vector approaches, see \citet{rodriguez2022word}).}

Figure \ref{fig:MLillustrate} provides an illustration of the model's structure. In particular, we model each \emph{character} as a vector (with each dimension representing some latent quality of that character), each \emph{word} as an ordered sequence of character vectors, and each \emph{organizational name} as an ordered sequence of word vectors learned from their character constituents. That is, first, we learn a good representation of words based on ordered characters. Then, we learn a good representation of organization names based on ordered words. Finally, we repeatedly optimize the system through backpropagation to minimize the loss function above.

Here, an important parameter is the dimension of the vector representation. We adopt a 1,024-dimensional numeric representation of organizational names, balancing computational efficiency with informational richness. 
Similar to other word embedding approaches, each dimension of the ultimate vector has a latent semantic value, although that value may be hard to interpret. 

\begin{figure}[t]
\begin{center}
\includegraphics[width=0.75\textwidth]{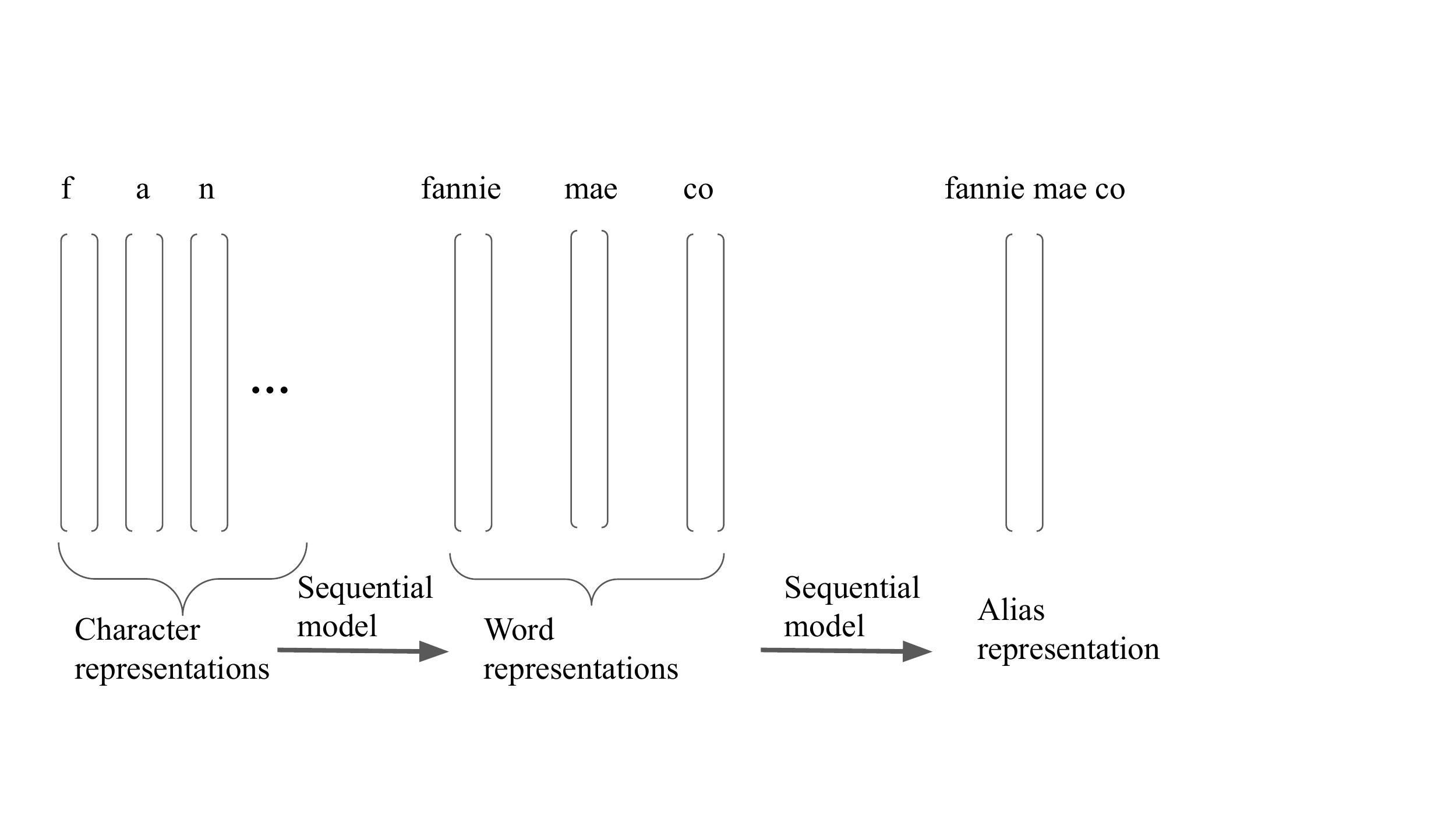}
\caption{
A high-level illustration of the multi-level neural network’s architecture. We learn from data how to represent, in a vector space, (a) the characters that constitute words and (b) the words that constitute organizational names. Each lower level is used to generate a higher-level representation in vector space via a flexible model, with better lower-level representations learned by tuning higher-order representations.
}\label{fig:MLillustrate}
\end{center}
\end{figure}

\begin{figure}[t]
\centering
\includegraphics[width=0.50\textwidth]{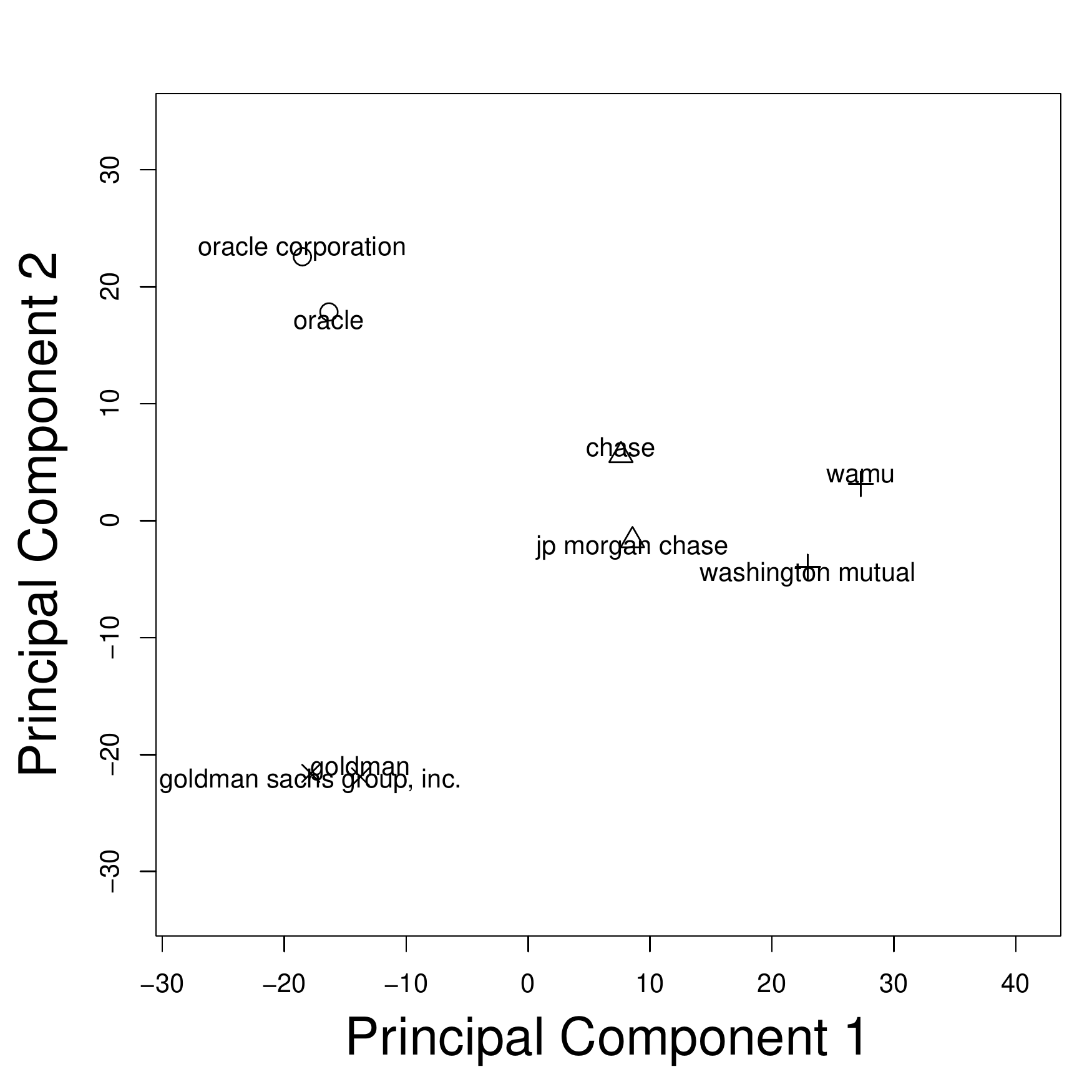}
\caption{
  Visualizing the machine learning output: Similar organizational names are close in this vector space, which has been projected to two dimensions using PCA.
  \label{fig:Embeddings}
}
\end{figure}

 In Figure \ref{fig:Embeddings}, we examine how the algorithm has mapped aliases into the embeddings space. Due to the difficulties of visualizing multi-dimensional data, we project the alias embedding space down to two dimensions via Principal Component Analysis (PCA). Pairs of aliases representing the same organization are represented by the same graphical mark type. We see that aliases representing the same organization are generally quite close in this embedding space. The model seems to be able to handle less salient information  well: ``oracle'' and ``oracle corporation'' are quite close in this embedding space even though the presence of the long word ``corporation'' would substantially affect string distance measures based only on the presence/absence of discrete letter combinations. While researchers may drop common words like ``corporation'' based on intuition, our optimized model learns which words to emphasize or ignore from data. 

While these examples are interesting and encouraging, they do not present a particularly rigorous test of the algorithm's performance. In Figure \ref{fig:showML}, we examine how well names that should match do match and how well names that should not match do not match according to the estimated model. In particular, we hold out from training 2,000 randomly chosen pairs of aliases sharing a URL (``matches'') and 2,000 randomly chosen pairs of aliases where a URL is not shared (``non-matches''). 
For the set of matches and non-matches, we provide density plots of the match quality under fuzzy matching and under our learning model.

The right panel of Figure \ref{fig:showML} considers predicted match probabilities for the out-of-sample set of match and non-match examples from the LinkedIn corpus. In particular, it shows density plots of the predicted probabilities of pairs that are matches and, separately, non-matches. If the algorithm is working as it should then the overall distribution of match probabilities for matches and non-matches should differ greatly. Indeed, this is what the figure finds. A KS test for assessing whether the probabilities are drawn from the same distribution yields a test statistic of 0.87 ($p < 10^{-16}$). A statistic of 0 indicates complete distribution overlap, while 1 signifies perfect separation. Encouragingly, we are closer to this second case. The left panel shows results with Jaccard-distance fuzzy matching, which yields KS test statistics ranging from 0.47 to 0.55, depending on the character $q$-grams used. 

\begin{figure}[t]
\includegraphics[width=0.50\textwidth]{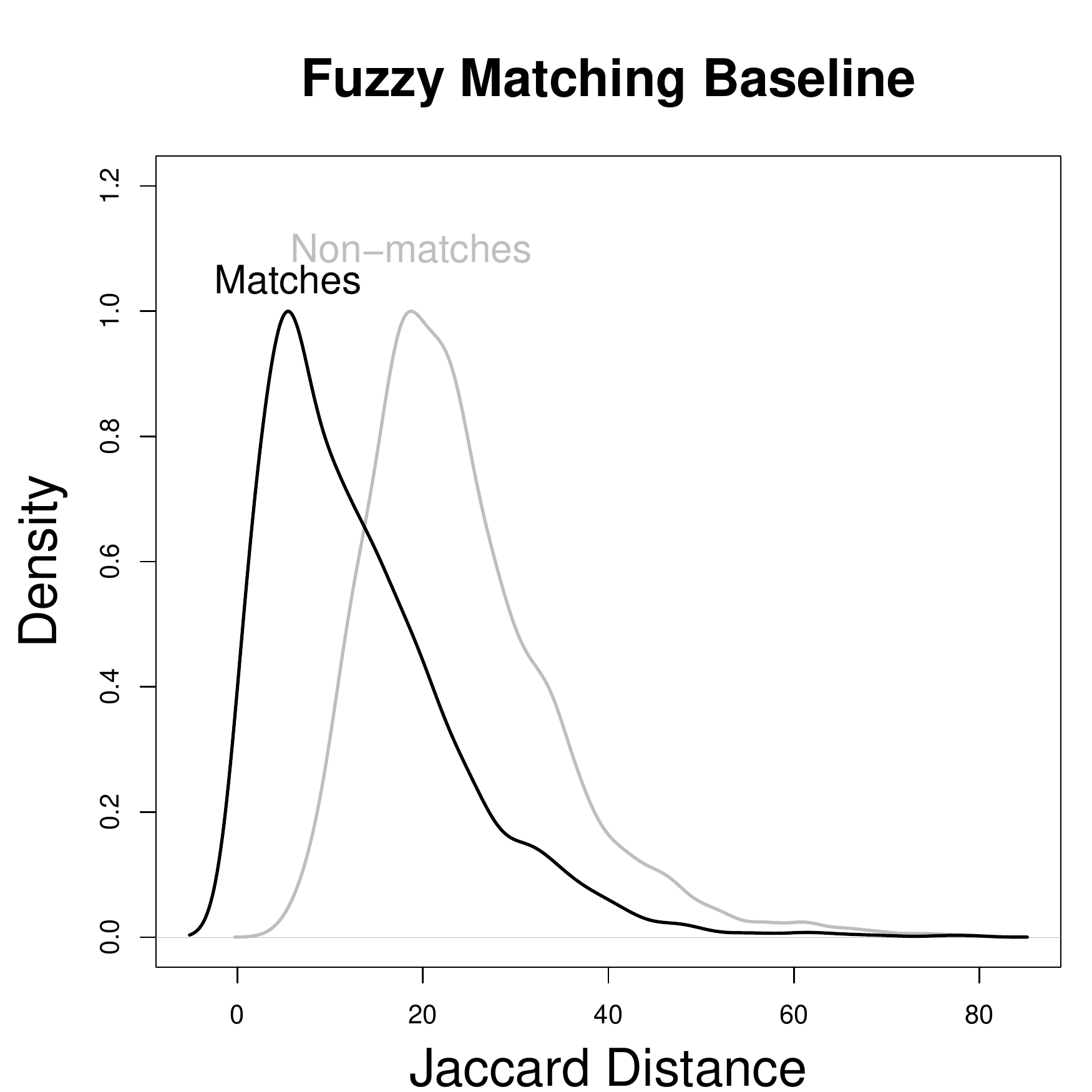}
\includegraphics[width=0.50\textwidth]{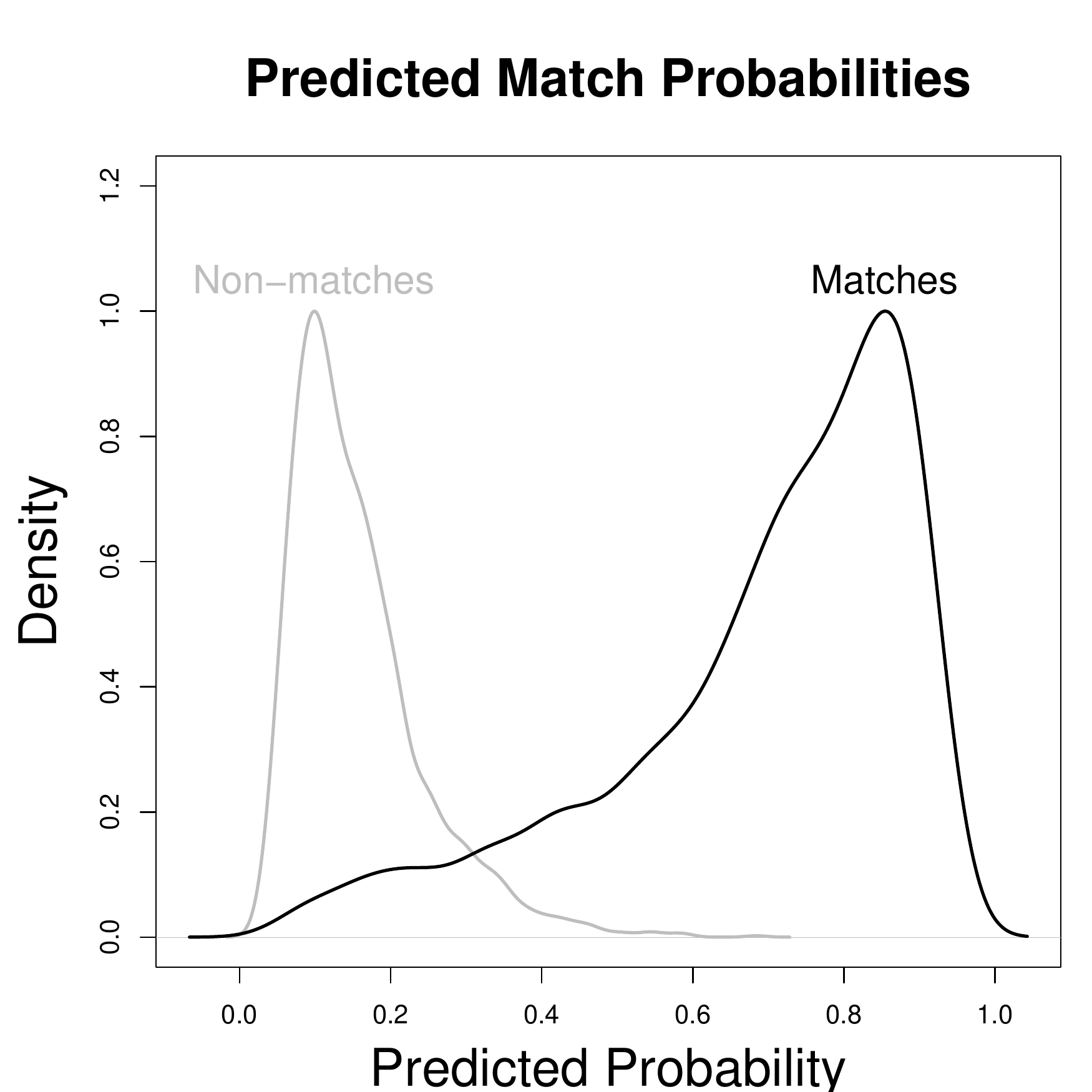}
\caption{Visualizing the machine learning model output: {\sc Left.} Fuzzy matching as a baseline generates distances between strings that have trouble distinguishing matches from non-matches in some cases {\sc Right.} On average, organizational alias matches have higher match probabilities compared with the set of non-matches.\label{fig:showML}}
\end{figure}

Despite these successes, there are true links that would remain hard to model using this prediction-oriented framework. For instance, the aliases ``Chase Bank'' and ``JP Morgan'' have a low match probability. To handle such cases, we next show how the LinkedIn data introduced in this paper can be used in a network-oriented approach to improve organizational record linkage. 

\subsection{Network-based Linkage Algorithms with LinkedIn Data}\label{s:Network}

As we have already seen, organizational names sometimes contain little semantic information; as a result, methods that focus on uncovering these meanings have a ceiling. Often, the relationship between two aliases for an organization is something that one simply has to know. The question then is how to best leverage the knowledge implicit in the LinkedIn network, bearing in mind that the raw data may not reveal the full depth of knowledge in the network. 

\begin{figure}
    \centering
    \includegraphics{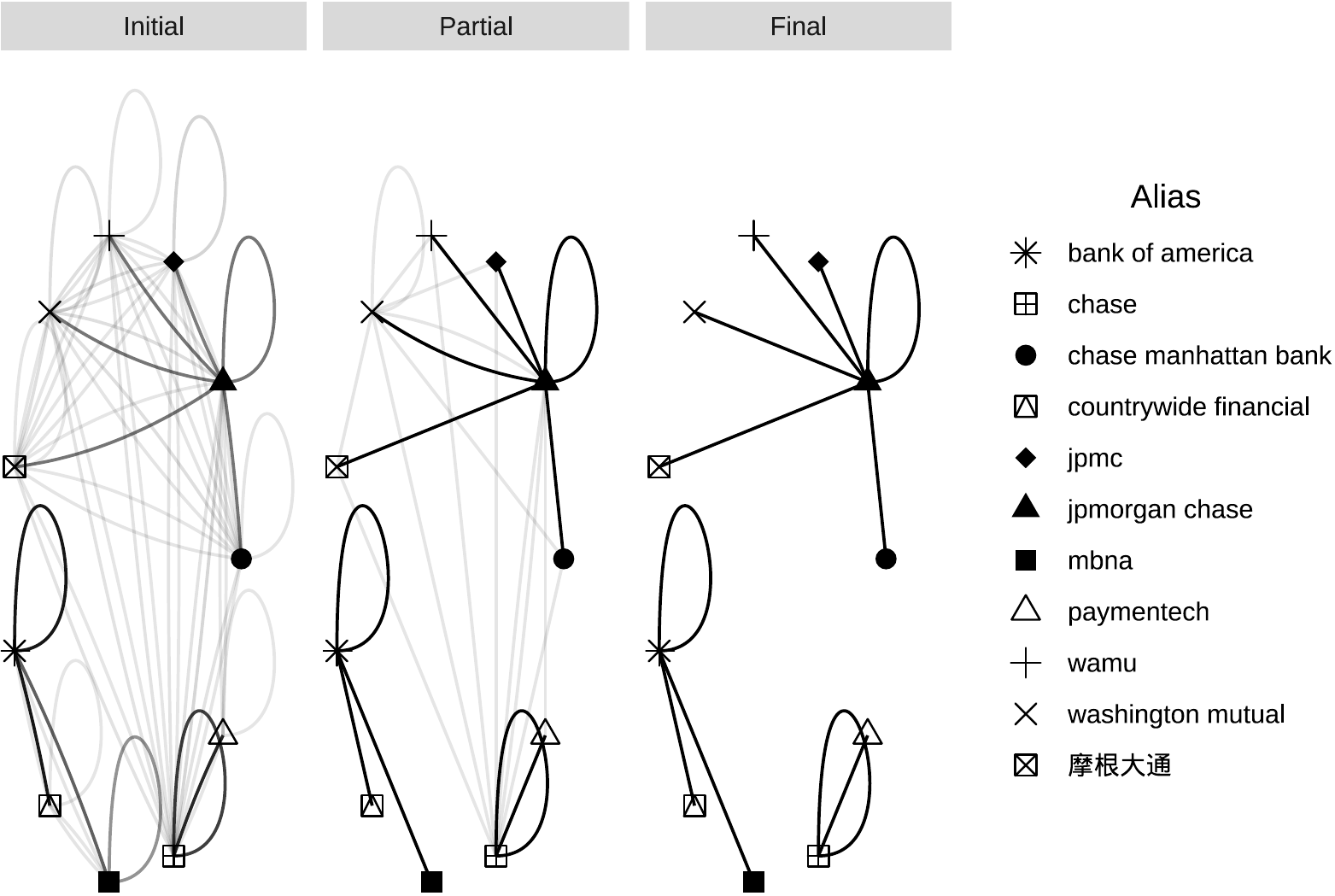}
    \caption{\textbf{LinkedIn Name Network and Community Detection Algorithm}. The figure illustrates how the community detection algorithm links organizational aliases for 11 aliases in the banking space. {\sc Left:} A weighted graph derived from raw counts of connections in the LinkedIn database. Due to some stronger connections between some nodes, the figure suggests some clusters visually, but there are also connections across clusters, which makes clustering a non-trivial problem. {\sc Middle:} The Markov clustering algorithm proceeds toward convergence. The connections between nodes are now stronger within and weaker across communities. A ``Bank of America'' community appears to have been detected. {\sc Right:} Community detection has converged to three clusters. The tight connection between JP Morgan and its subsidiaries contrasts with what is found through word embeddings, where these names are rather distant in the machine learning-based vector space (compare the positioning of some of the same aliases in Figure \ref{fig:Embeddings}).}
    \label{fig:clustering}
\end{figure}

Instead of viewing the record linkage task as matching two lists of organization names directly, one can instead view it as connecting these names on a graph. A tricky point, of course, is that the names of the organizations in one's lists may not actually be on the graph. We address this issue below when thinking about an ensemble strategy (and also through our third application). But even assuming the names are on the graph, one must consider the sense of connectedness that is most useful.

The simplest concept of connectedness would assert that two aliases are linked if someone has attributed the same URL to both names. This notion would often fail to follow transitivity. In other words, if $A$ and $B$ refer to the same organization and $B$ and $C$ refer to the same organization, then $A$ and $C$ should refer to the same organization---but, despite this, they may not share a URL. So, without care, we may miss this connection. It is tempting then to insist on transitivity in alias names. Implicitly, doing so casts the problem of record linkage as placing an organization in a particular connected component of the LinkedIn network. An issue here is that, if there are spurious links, then many components will be merged where there is little evidence to support such an action. An approach that is in between, allowing for \textit{some} transitivity when the evidence is sufficiently strong but not when the evidence is weak, is desirable. Community detection methods aim to find this sweet spot.

Because community detection is a well-studied problem that occurs in many applied contexts \citep{rohe2011spectral}, we consider two algorithms established in the methodological literature: Markov clustering  \citep{van2008graph} and greedy clustering \citep{clauset2004finding}. We focus on these algorithms because they model the network in distinct ways and are computationally efficient. Implementation details are in Appendix II; here, we offer a brief sketch of each.

Figure \ref{fig:clustering} shows how we can represent the data source explicitly as a network for Markov clustering. Here, 11 organizational aliases are presented as nodes. Aliases are connected by edges. In principle, these could be directed or undirected, weighted or unweighted depending on one's modeling strategy. The figure shows edges with weights that follow a naive Bayesian strategy for calibrating the amount of information between names (we use a similar probability calculation as in Eq. \ref{eq:bayes}). Under this approach, the probability that $A$ and $B$ is connected is the same as the probability that $B$ and $A$ are connected. Therefore, this yields a weighted, undirected graph. Notable, we see the strong ties where the connection is surprising based on semantic information, such as ``chase'' and ``washington mutual.'' Visually, it is clear that there are two to three clusters where links are denser, but there are also occasional ties across the clusters that {\it ex ante} are hard to identify as spurious or real. These clusters of nodes with relatively dense connections are the ``community'' of aliases we wish to discover. 

Markov clustering applies arithmetic operations to the edges of the graph that alternately diminish weak links in the graph and enhance strong ones. The middle panel of Figure \ref{fig:clustering} shows the partial completion of this algorithm while the right panel shows it at convergence, where each organization is placed in a single ``community'' identified by the alias most prominent in it (for example, ``jp morgan chase'' and ``bank of america'').

In contrast to Markov clustering, greedy clustering is an iterative algorithm that begins by assuming each node is its own community and then merges communities that would result in the largest increase in the overall ``quality'' of the network structure. We use one of the most ubiquitous quality measures called a modularity score. This score is \(0\) when community ties between aliases and URLs occur as if communities were assigned randomly. It gets larger when the proposed community structure places aliases that tend to link to the same URLs in the same community \citep{clauset2004finding}. While the Markov clustering algorithm requires edges to have probability weights, greedy clustering does not, which enables community detection with a bipartite (as opposed to adjacency) representation. 

Ultimately, we find somewhat better performance with this bipartite representation of the LinkedIn network where the names and URLs are both considered nodes and the links only occur between names and URLs if there is an attribution in the LinkedIn database (with edge weights given by the number of times two attributions are made).

\subsection{Joint Network and Prediction-based Record Linkage Using the LinkedIn Corpus}\label{s:UnifiedApproach}

The LinkedIn-calibrated machine learning model uses complex semantic information to assist matching but does not make use of graph-theoretic information. The network-based methods use network information but do not use semantic information to help link names in that network. To get the best of both worlds, we propose a third, unified approach that uses both the semantic content and graph structure of the LinkedIn corpus. 

\begin{figure}
\centering
\captionsetup[sub]{              
    labelformat=step}        
\begin{subfigure}[b]{0.9\textwidth}
   \includegraphics[width=1\linewidth]{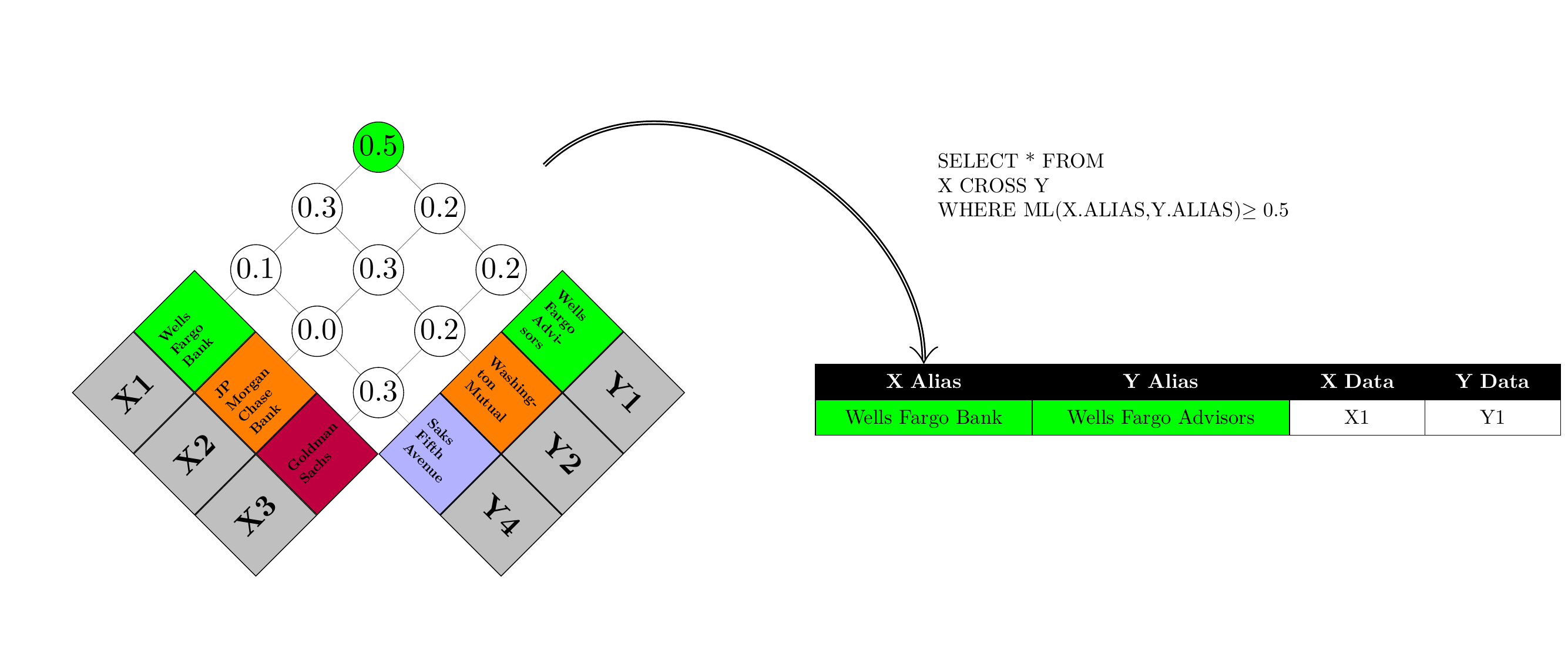}
   \caption{Direct linkage through machine learning-optimized string matching.}
   \label{fig:Unified-a} 
\end{subfigure}
\begin{subfigure}[b]{0.9\textwidth}
   \includegraphics[width=1\linewidth]{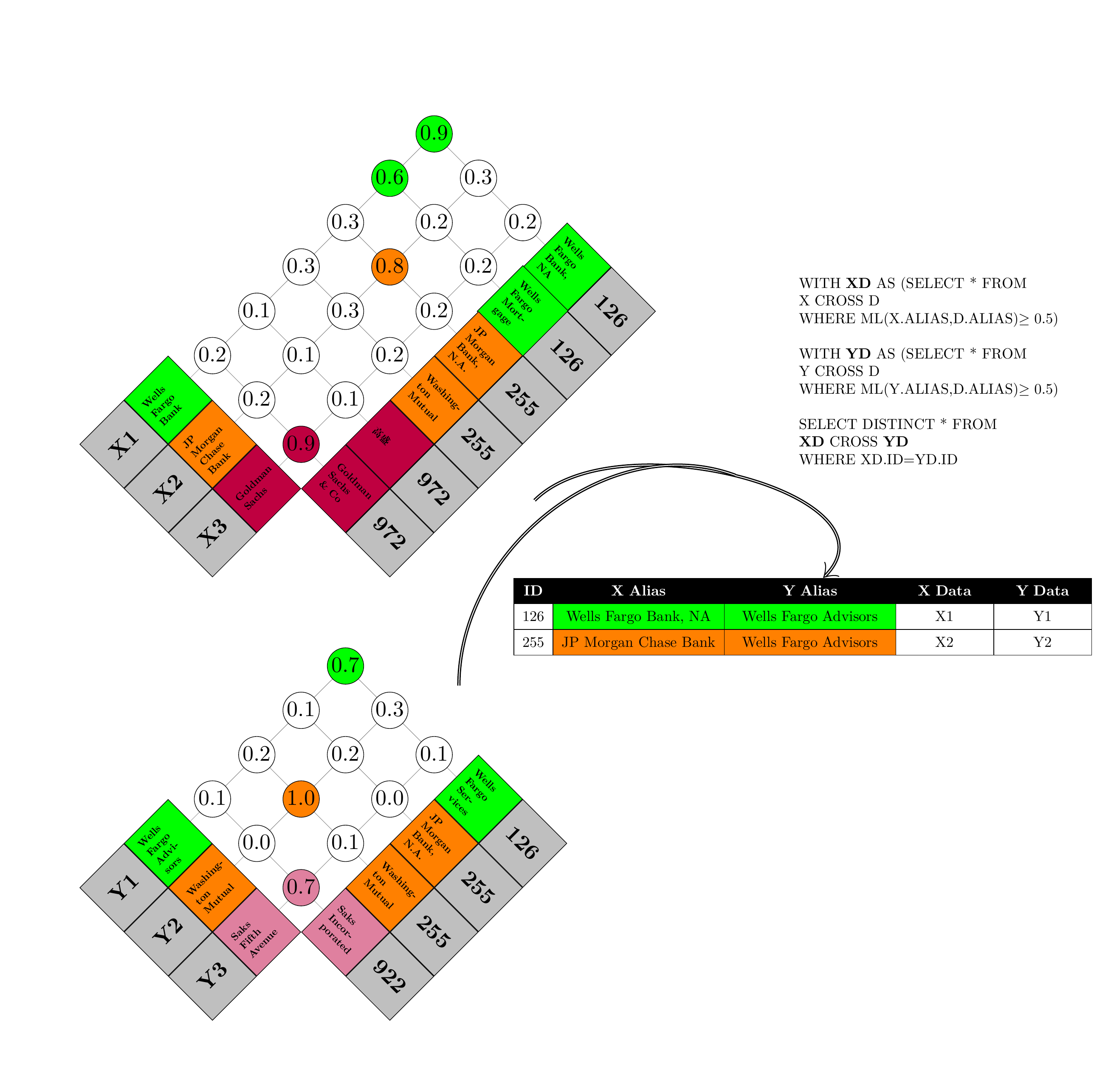}
   \caption{Indirect linkage through a directory constructed using community detection.}
   \label{fig:Unified-b}
\end{subfigure}
\caption[Checkered Flags]{Checkered flag diagrams illustrating a unified approach to name record linkage using the LinkedIn corpus.}
\label{fig:Unified}
\end{figure}

The unified approach is an ensemble of both network and machine learning methods and involves three steps. Figure \ref{fig:Unified} returns to the example at the beginning of this section involving the merge of a dataset about Wells Fargo Bank, JP Morgan Chase Bank, and Goldman Sachs ($\bX$) with another dataset about Wells Fargo Advisors, Washington Mutual, and Saks Fifth Avenue ($\bY$). The figure presents several checkered flags illustrating the multi-step approach. In step (a), machine learning-assisted name linkage is directly applied between the two datasets. Similar to fuzzy string matching, scores are calculated on the cross product of two sets of names; scores exceeding a threshold (set to 0.5 in the figure) are said to match. In this particular example, fuzzy matching could produce similar results, but the thresholds and scores would differ, and, as a result, so too would performance. In step (b), machine learning-assisted name linkage is applied to an intermediary directory built using community detection. We attempt to place $\bX$ in their proper community and $\bY$ in their proper community, and then we consider entries in $\bX$ and $\bY$ as linked if they are placed in the same community.

This example shows how the unified approach can put to use the potential of both methods presented thus far. Through step (a), it can link datasets for organizations that do not appear on LinkedIn but whose naming conventions are similar. Through step (b), the unified method picks up on relationships that are not apparent based on names and require specialized domain expertise to know. We now turn to the task of assessing how these different algorithmic approaches and representations of the LinkedIn corpus perform in examples from contemporary social science. 

\section{Evaluation Tasks}\label{s:Illustration}

\subsection{Method Evaluation}\label{method-evaluation}
Before we describe the illustrative tasks, we first introduce our comparative baseline and evaluation metrics. This introduction will help put the performance of the methods into context. 

\subsubsection{Fuzzy String Matching Baseline}\label{fuzzy-string-matching-baseline}

We examine the performance of the LinkedIn-assisted methods against a fuzzy string-matching baseline. While there are many ways to calculate string similarity, we continue to focus on fuzzy string matching using the Jaccard distance measure to keep the number of comparisons manageable. Other string discrepancy measures, such as cosine distance or edit distance, produce similar results. 

\subsubsection{A Machine Learning Baseline}\label{ML-baseline}

We also examine the performance of the LinkedIn-assisted methods against a machine learning baseline, ``DeezyMatch,'' that uses a recurrent neural network-based fuzzy matching approach outlined in \citet{hosseini2020deezymatch}, with hyperparameters left at their defaults. This approach will provide a helpful baseline for contextualizing performance. 

\subsubsection{A Network Approach Baseline}\label{Network-baseline}

We also examine performance against a simple method (hereafter, ``lookup'') that uses the LinkedIn data as a giant lookup table for organizations to assess the relative value-added of the clustering algorithms. In this approach, we consider two aliases as matched if they link to the same URL at least once in the LinkedIn corpus. 

\subsubsection{Performance Metrics}\label{performance-metrics}

We consider two measures of performance. First, we consider the fraction of true matches discovered as we vary the acceptance threshold. This value is defined to be 
\begin{equation} 
\textrm{True positive rate} = \frac{\textrm{\# of true positives found}}{\textrm{\# of true positives in total}} 
\end{equation} 
This measure is relevant because, in some cases, researchers may be able to manually evaluate the set of proposed matches, rejecting false positive matches. The true positive rate is therefore more relevant for the setting in which scholars use an automated method as an initial processing step and then evaluate the resulting matches themselves, as may occur for smaller match tasks.

While the true positive rate captures our ability to find true matches, it does not weigh the cost involved in deciding between true positives and false positives (i.e., matches the algorithm finds that are not, in fact, real). Failure to consider the costs of false positives can lead to undesirable conclusions about the performance of algorithms. ``Everything matches everything'' is a situation that ensures all true matches are found, but the results are not informative. Given such concerns, we also examine a measure that considers the presence of true positives, false positives, and false negatives known as the \(F_\beta\) score, defined as 
 \begin{equation} 
 F_\beta = \frac {(1 + \beta^2) \cdot \mathrm{true\ positive} }{(1 + \beta^2) \cdot \mathrm{true\ positive} + \beta^2 \cdot \mathrm{false\ negative} + \mathrm{false\ positive}}, 
 \end{equation}
where the $\beta$ parameter controls the relative cost of false negatives compared to false positives \citep{lever2016classification}. In the matching context, errors of \emph{inclusion} are typically less costly than errors of \emph{exclusion}: the list of successful matches is easier to double-check than the list of non-matched pairs. For this reason, we examine the \(F_2\) score, a choice used in other evaluation tasks (e.g.,  \citet{devarriya2020unbalanced}), which weighs false negatives more strongly than false positives. 

\subsubsection{Comparing Algorithm Performance Across Acceptance
Thresholds}\label{comparing-algorithm-performance-across-acceptance-thresholds.}
Approximate matching algorithms have a parameter that controls how close a match must be to be acceptable. \footnote{Results for the network-based linkage approaches also vary with this parameter because we first match aliases with entries in the directory in order to find the position of those aliases within the community structure of LinkedIn.} Two algorithms might perform differently depending on how the acceptance threshold parameter is set. This threshold is not directly comparable across algorithms. For instance, a change of 0.1 in the match probability tolerance under the ML algorithm implies a much different change in matched dataset size than a 0.1 change in the Jaccard string distance tolerance. To compare the performance of these algorithms, our figures and discussion focus on the size of matched datasets induced by an acceptance threshold. The most stringent choice produces the smallest dataset (i.e., consisting of the exact matches), while the lowest possible acceptance threshold produces the cross-product of the two datasets (i.e., everything matches everything). Between the two, different thresholds produce datasets of different sizes. By comparing performance across matched dataset sizes, we can evaluate how the algorithms perform for different acceptance thresholds.

\subsection{Task 1: Matching Performance on a Lobbying
Dataset}\label{task-1-matching-performance-on-a-lobbying-dataset}

We first illustrate the use of the organizational directory on a record linkage task involving lobbying and the stock market. \citet{Libgober2020a} shows that firms that meet with regulators tend to receive positive returns in the stock market after the regulator announces the policies for which those firms lobbied. These returns are significantly higher than the positive returns experienced by market competitors and firms that send regulators written correspondence. Matching meeting logs to stock market tickers is burdensome because there are almost 700 distinct organization names described in the lobbying records and around 7,000 public companies listed on major US exchanges. Manual matching typically involves research on these 700 entities using tools such as Google Finance. While the burden of researching 700 organizations in this fashion is not enormous,  \citet{Libgober2020a} only considers meetings with one regulator. If one were to increase the scope to cover more agencies or all lobbying efforts in Congress, the burden could become insurmountable.

Treating the human-coded matches in \citet{Libgober2020a} as ground truth, results show how the incorporation of the LinkedIn corpus into the matching process can improve performance. Figure \ref{fig:performance} shows that the LinkedIn-assisted approaches almost always yield higher \(F_2\) scores and true positives across the range of acceptance thresholds. The highest \(F_2\) score is over 0.6, which is achieved through the unified approaches, the machine learning approach, and the bipartite graph-assisted matching. The best-performing algorithm across the range of acceptance thresholds is the unified approach using the bipartite network representation when combined with the distance measure obtained via machine learning. The percentage gain in performance of the LinkedIn-based approaches is higher when the acceptance threshold is closer to 0; as we increase the threshold so that the matched dataset is ten or more times larger than the true matched dataset, the \(F_2\) score for all algorithms approaches 0, and the true positive rate approaches 1. 

\begin{figure}[t]
\includegraphics[width=0.35\textwidth]{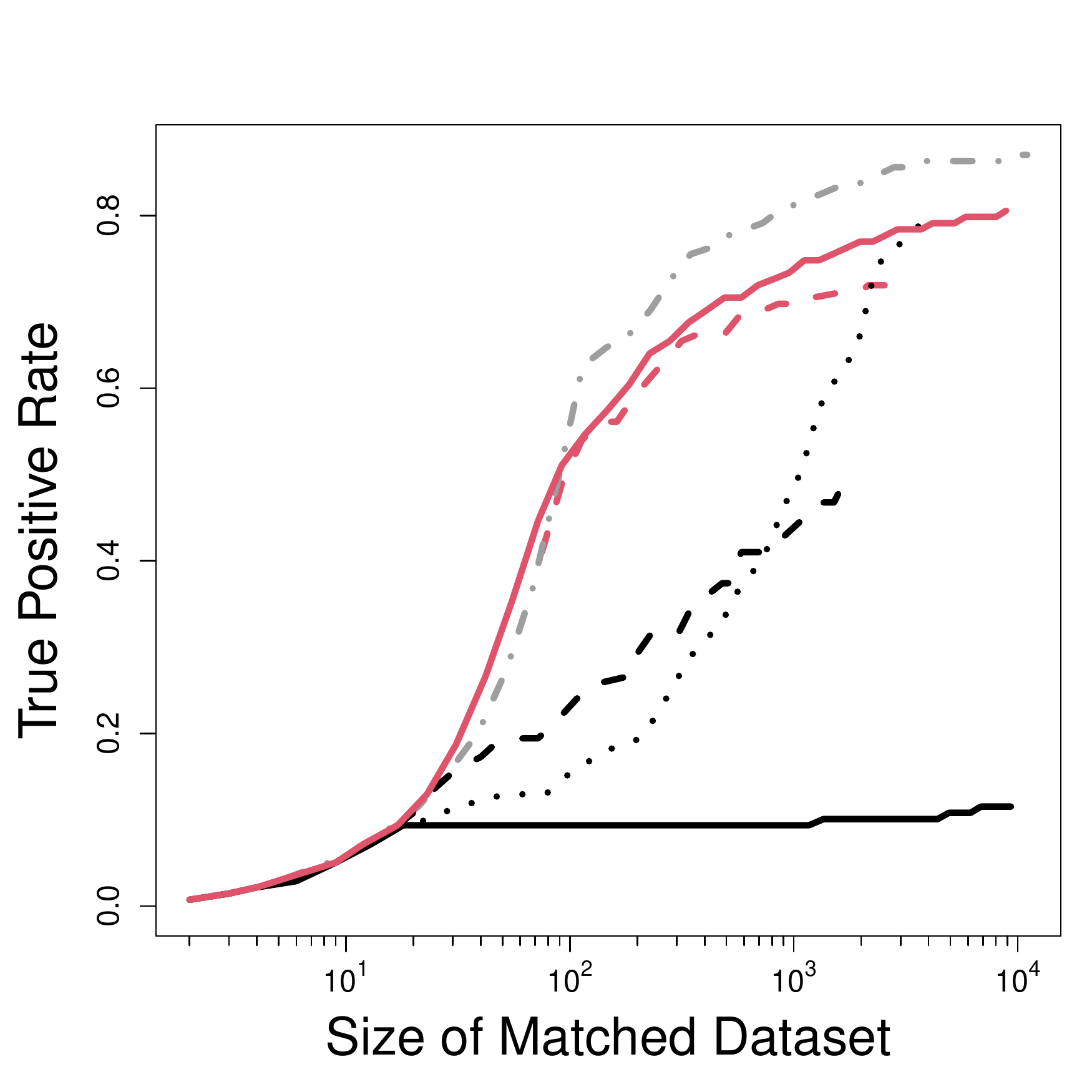}\includegraphics[width=0.35\textwidth]{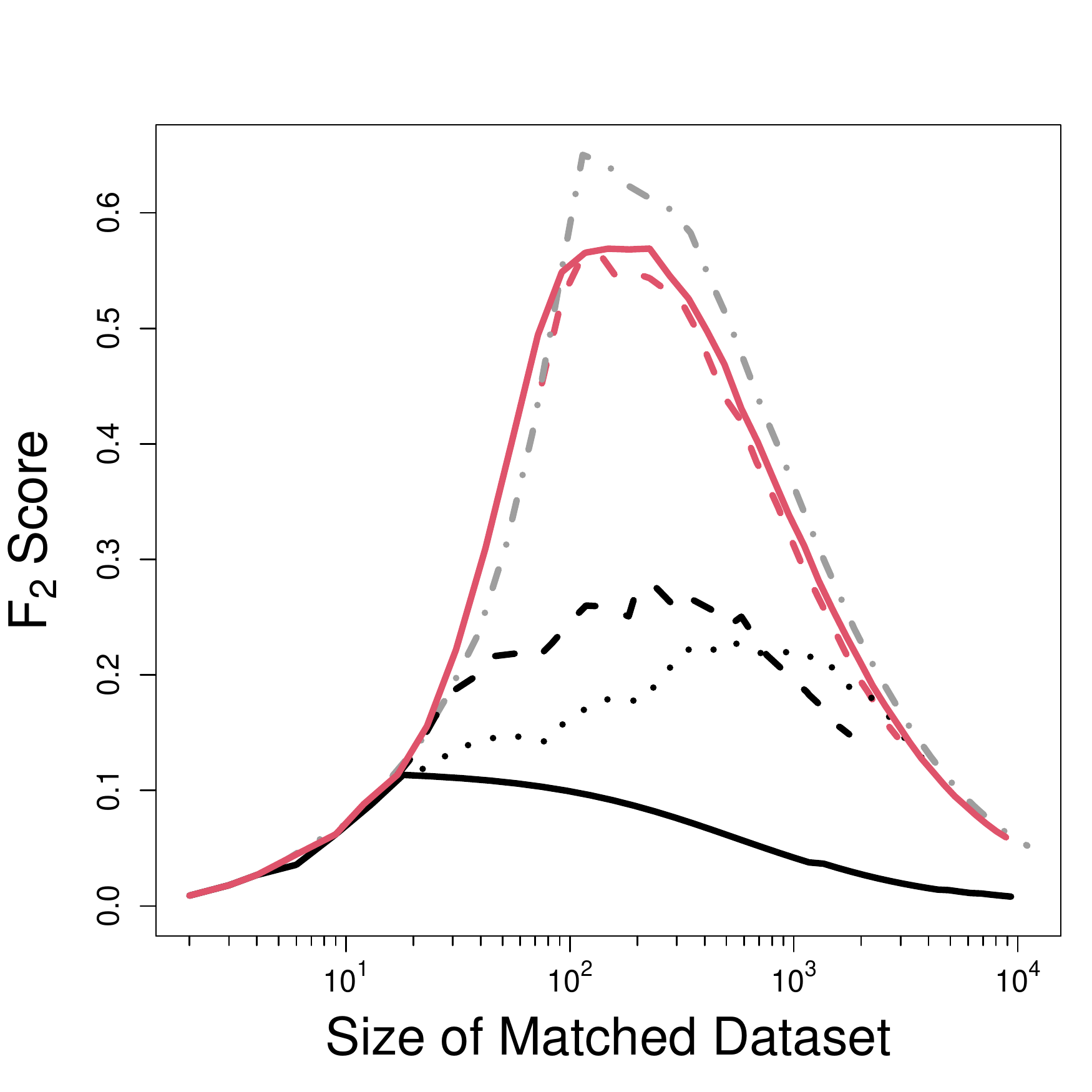}
\includegraphics[width=0.2\textwidth]{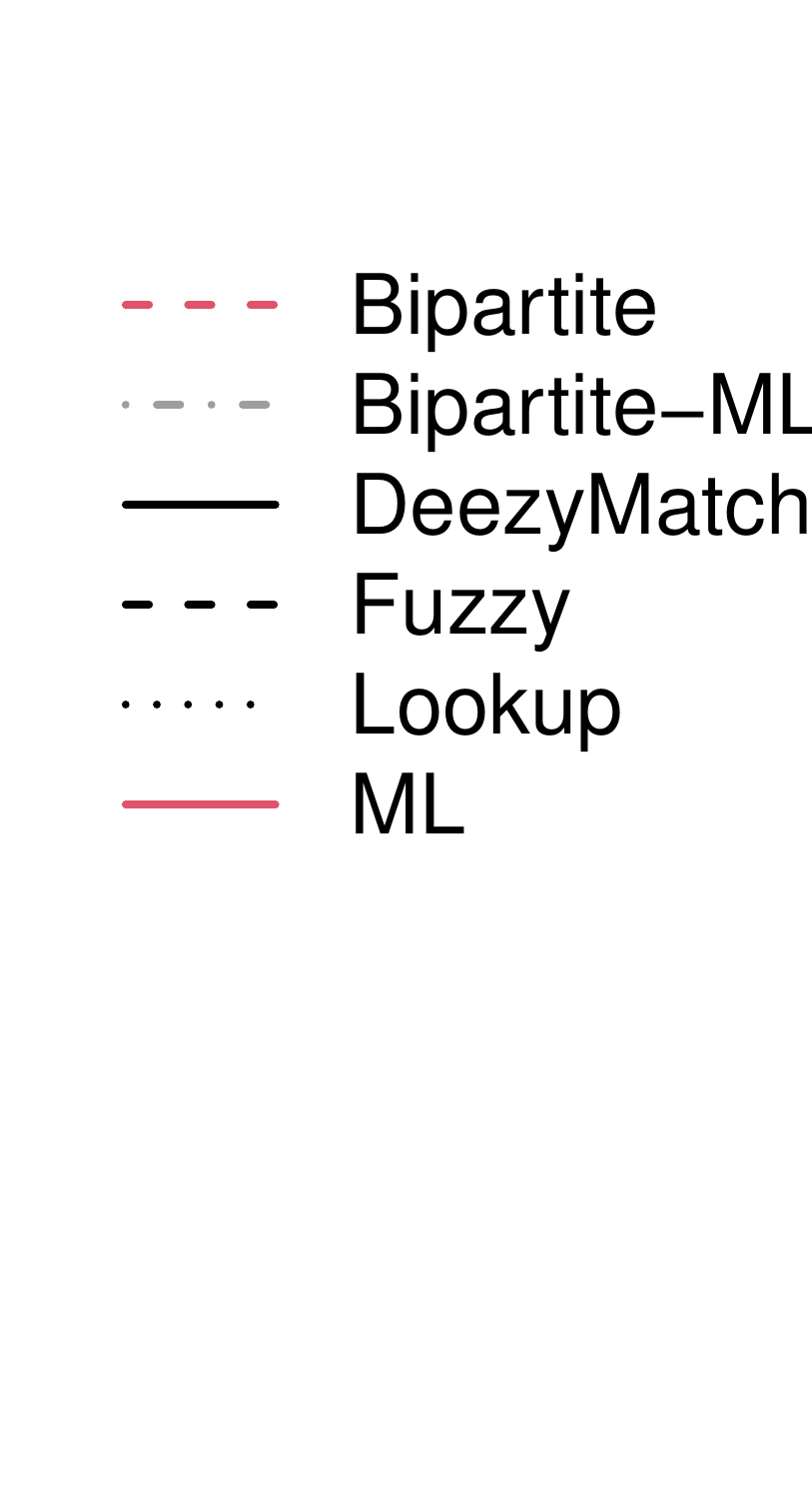}
\caption{We find that dataset linkage using any one of the approaches with the LinkedIn network obtains favorable performance relative to fuzzy string matching both when examining only the raw percentage of correct matches obtained {\sc (Left Panel)} and when adjusting for the rate of false positives and false negatives in the $F_2$ score {\sc (Right Panel)}. In both figures, higher values along the $Y$-axis are better. 
The ``Bipartite'' refers to the Bipartite network-based approaches to linkage. ``ML'' refers to the machine learning approach introduced above. ``Fuzzy,'' ``DeezyMatch,'' and ``Lookup'' refer to the string distance, machine learning, and network baselines. ``Bipartite-ML'' refer to the ensemble of ``Bipartite'' and ``ML.'' See Figure \ref{fig:performanceAppendix} for full results with Markov approaches included. 
\label{fig:performance}}
\end{figure}

For illustration, let us examine a case where it successfully identifies a correct match that fuzzy matching fails to detect. Fuzzy matching fails to link the organizational log entry associated with ``HSBC Holdings PLC'' to the stock market data associated with ``HSBC.'' Their fuzzy string distance is 0.57, which is much higher than the distance of ``HSBC Holdings PLC'' to its fuzzy match (0.13 for ``AMC Entertainment Holdings, Inc.''). ``HSBC Holdings PLC,'' however, has an exact match in the LinkedIn-based directory, so the two organizations are successfully paired using the patterns learned from the LinkedIn corpus.

Another relevant consideration in applied matching tasks is compute time. In Section \ref{s:Time}, we document the runtime of each approach in the different applications. As expected, the network-based approaches have the greatest computational cost, as some measure of distance between each candidate observation must be computed against all of the hundreds of thousands of entities in the LinkedIn corpus. For these network approaches, the runtime is on the order of several hours for this roughly 700 by 7,000 name merge. By contrast, fuzzy matching runs in less than 1 minute; the machine learning approach without the combined network approach runs in roughly 5 minutes on 2024 hardware. Scaling the best methods is, therefore, a potential concern as one reaches datasets with organizations numbering in the tens or hundreds of thousands. Back-of-the-envelope calculations suggest that a 10,000 by 10,000 organization match would potentially have a 2-3 day runtime using full Bipartite-ML, which is long but not unacceptable, as is it performed once in the course of an entire project without much researcher intervention. 

Additional strategies would likely be necessary to scale to a 100,000 by 100,000 name-matching problem, as the best performing, but slowest, algorithm would run somewhere around 255 days, a wait time not realistic for the research iteration process. Two such strategies are parallelization and locality-sensitive hashing. While parallelization can speed up easily subdivided problems like name matching, most computational costs are spent checking pairs that have low match probabilities. Techniques such as locality-sensitive hashing can improve speed by avoiding  comparisons between unlikely matches \parencite{green2023zoomerjoin}.

\begin{table}[!htbp] \centering 
  \caption{Run time on the meetings data analysis.} 
  \label{tab:ExTime2} 
\begin{tabular}{@{\extracolsep{5pt}} cc} 
\\[-1.8ex]\hline 
\hline \\[-1.8ex] 
Algorithm & Run Time (mins) \\ 
\hline \\[-1.8ex] 
Bipartite & 13.12 \\ 
Bipartite-ML & 251.38 \\ 
DeezyMatch & 0.24 \\ 
Fuzzy & 0.27 \\ 
Lookup & 1.35 \\ 
Markov & 8.61 \\ 
Markov-ML & 113.00 \\ 
ML & 1.63 \\ 
\hline \\[-1.8ex] 
\end{tabular} 
\end{table}

Overall, the results from this task illustrate how the LinkedIn-assisted methods appear to yield better performance than commonly used alternative methods, such as fuzzy matching, in the ubiquitous use case when researchers do not have access to shared covariates across organizational datasets. 

\subsection{Task 2: Linking Financial Returns and Lobbying Expenditures from Fortune 1000 Companies}\label{task-3-linking-financial-returns-and-campaign-contributions-from-fortune-1000-companies}

In the next evaluation exercise, we focus on a substantive question drawn from the study of organizational lobbying: do bigger companies lobby more? Prior research \citep{chen2015corporate} leads us to expect a positive association between company size and lobbying activity: larger firms have more resources that they can use in lobbying, perhaps further increasing their performance \citep{ridge2017beyond,eun2021aspirations}. Our reason for focusing on an application where there are \emph{such} strong theoretical expectations is to illustrate how results from different organizational matching algorithms can influence one's findings---something that would not be possible without well-established theory about what researchers should find.

For this exercise, we use firm-level data on the total dollar amount spent between 2013-2018 on lobbying activity. This data has been collected by Open Secrets, a non-profit organization focused on improving access to publicly available federal campaign contributions and lobbying data \citep{OpenSecrets}. We match this firm-level data to the Fortune 1000 dataset on the largest 1,000 US companies, where the measure of firm size we focus on is the average total assets in the 2013-2018 period. The key linkage variable will be organizational names that are present in the two datasets, that is to say, the name of the company according to Fortune and according to OpenSecrets. We manually obtained hand-coded matches to provide ground truth data.

In Figure \ref{fig:fortuneSubstantive}, we explore the substantive implications of different matching choices---how researchers' conclusions may be affected by the quality of organizational matches. We see that the coefficient relating log organizational assets to log lobbying expenditures using the human-matched data is about 2.5. In the dataset constructed using fuzzy matching, this coefficient is underestimated by about half. The situation is better for the datasets constructed using the LinkedIn-assisted approaches, with the effect estimates being closer to the true value. For all algorithms examined in Figure \ref{fig:fortuneSubstantive}, there is significant attenuation bias towards 0 in the estimated coefficient as we increase the size of the matched dataset, as poor-quality matches inject noise into estimation. Overall, we see from the right panel that match quality depends on algorithm choice as well as string distance threshold, with the LinkedIn-based approaches capable of estimating a coefficient within the 95\% confidence bounds of the ground truth estimate. Fuzzy matching and DeezyMatch, at their best, find an estimate that is only half as large in magnitude as the true value. 

\begin{figure}[t]
\centering 
\includegraphics[width=0.5\textwidth]{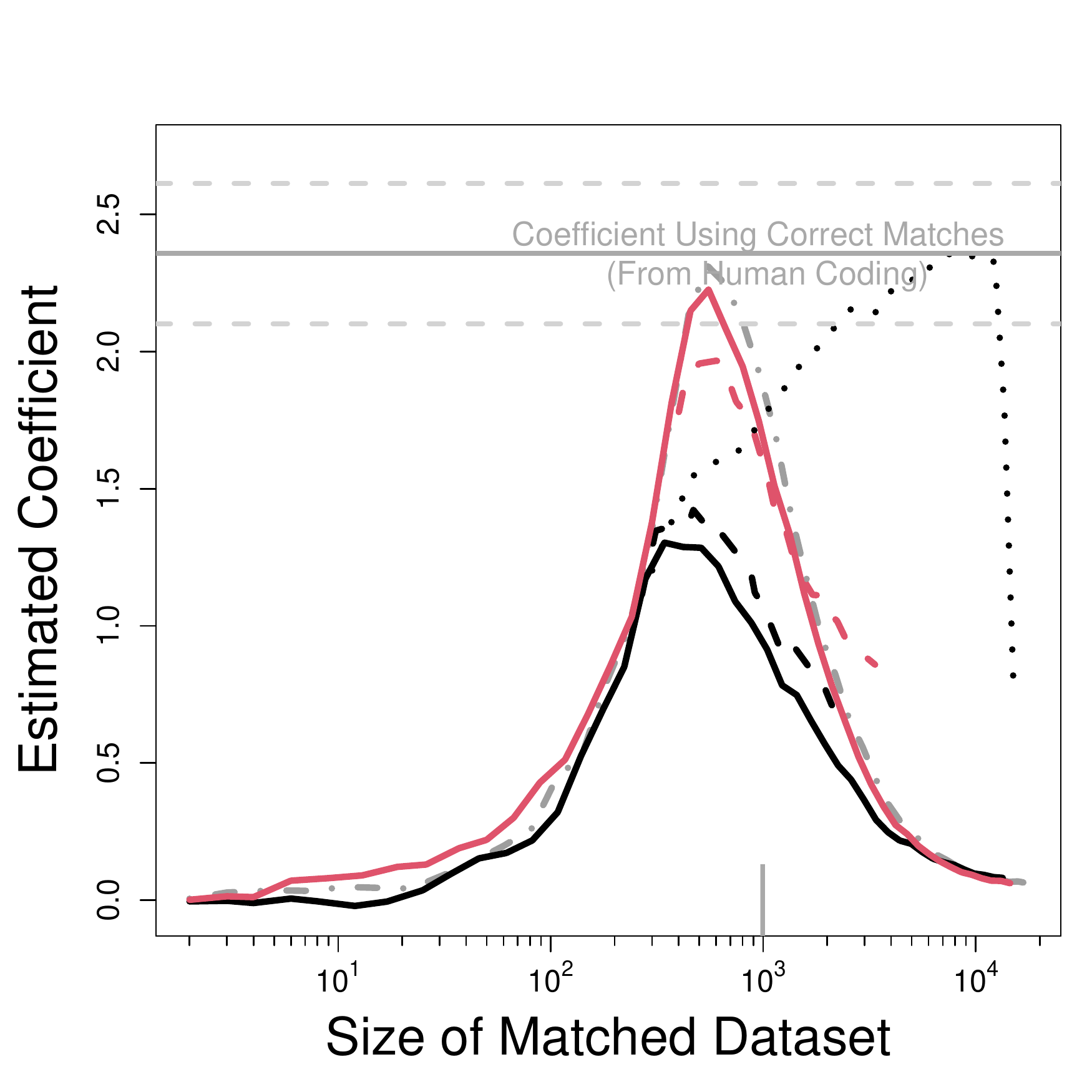}
\includegraphics[width=0.3\textwidth]{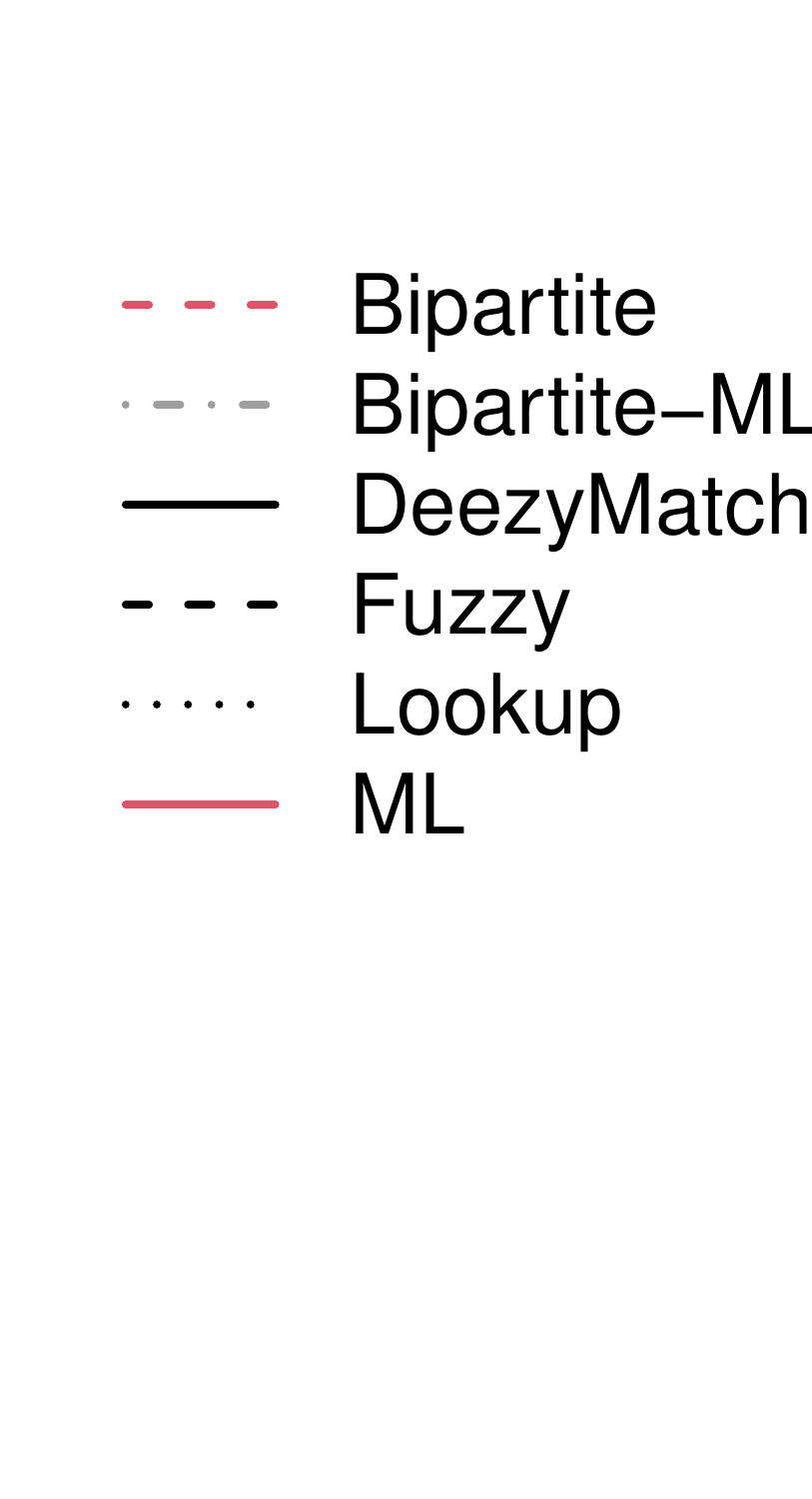}
\caption{The coefficient of log(Assets) for predicting log(1+Expenditures) using the ground truth data is about 2.5 (depicted by a bold gray line; 95\% confidence interval is displayed using dotted gray lines). At its best point, fuzzy matching underestimates this quantity by about half. The LinkedIn-based matching algorithms recover the coefficient better. See Figure \ref{fig:fortuneSubstantiveAppendix} for full results with Markov approaches included. 
 \label{fig:fortuneSubstantive}}
\end{figure}

\subsection{Task 3: Out-of-Sample Considerations: Merging YCombinator \& PPP Data}\label{s:YCombinator}

A final question is about how these methods perform in ``out-of-sample'' testing, which is performed with organizations that we do not expect to be well-represented in the current version of the LinkedIn data for whatever reason. While our methods could be adapted to use more recent versions of the LinkedIn data, LinkedIn data from 2017 cannot directly describe organizations that did not exist then. In this task, we analyze data from the period after the main data collection took place in an effort to understand the strengths and limitations of the various linkage strategies in this context. 

Here, we first examine data from a YCombinator directory on incubator startups. The dataset contains a collection of startups involved in the YCombinator seed-funding program, detailing their name, website, business model, team size, and development stage. This data provides a snapshot of the companies' early progression, from inception to public trading or acquisition. The YCombinator program was launched in 2005; for a purer out-of-sample test, we subset the data to the 2017-2024 period.

We merge these startups to the Paycheck Protection Program (PPP) loan database. The PPP (2020-2021) was a program aimed at providing financial relief to businesses during the COVID-19 pandemic. The dataset includes entries with key financial metrics, such as loan amount, approval date, borrower name and address, and employment impact. The task of matching startups to the PPP loan data could be relevant for evaluating the role of these loans on long-term firm survival as well as for thinking about the regulatory advantages that come from affiliation with a business network like YCombinator. Importantly, none of these covariates overlap with the Y-Combinator data. We subset both sets of data to target businesses within the San Francisco area. 

As expected, we find in Figure \ref{fig:YCombFig} that the linkage approaches only using the directory of firms with established LinkedIn pages as of 2017 provide no gain in the overall $F_2$ score relative to fuzzy matching. If these methods were rebuilt with a subsequent scrape of the LinkedIn database, they would likely do better with these new organizations. That said, the machine learning approach still provides a boost over fuzzy matching: the approach has inferred more enduring information about the link probability between companies based on the semantic content of names.  

\begin{figure}[t]
\begin{center}
\includegraphics[width=0.5\textwidth]{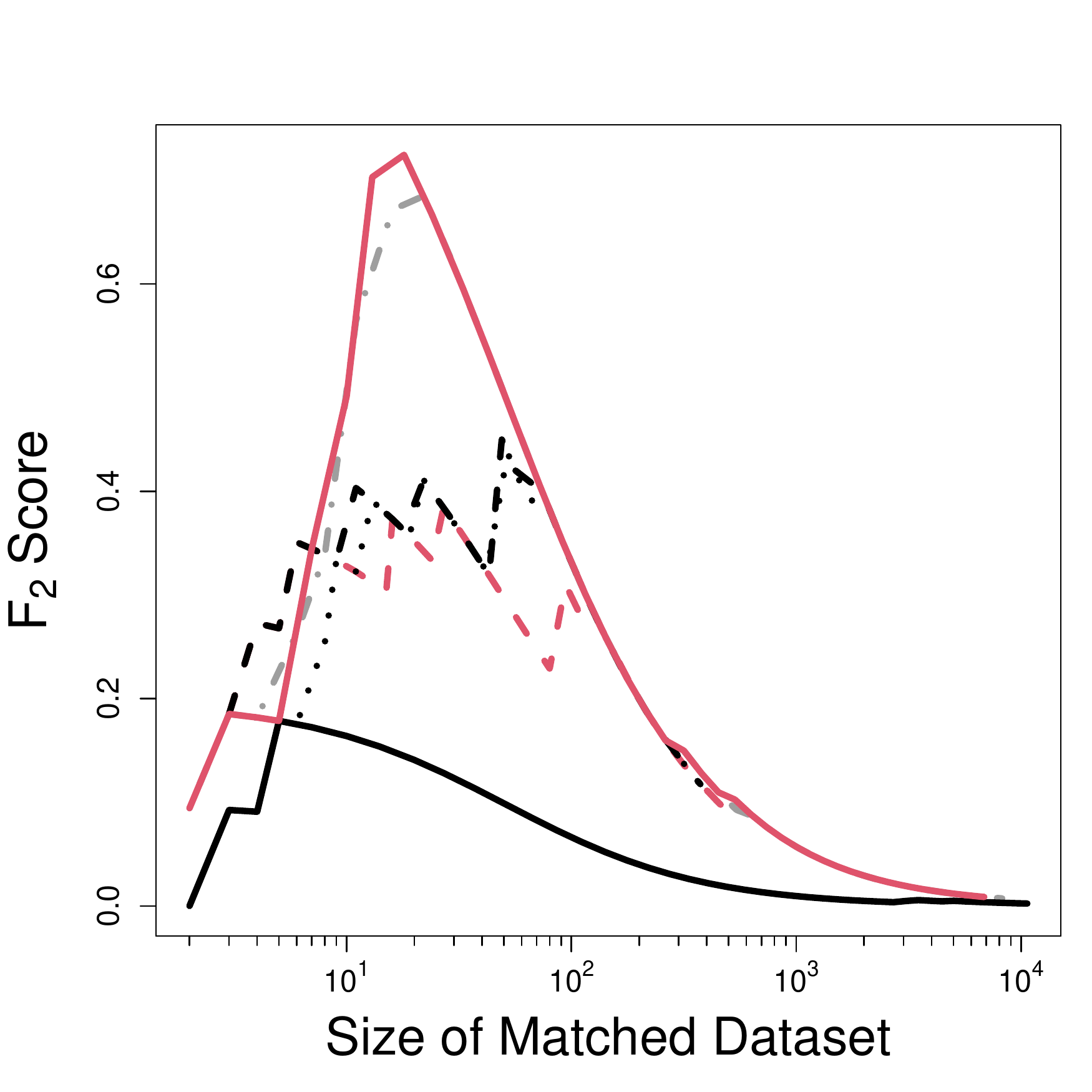}
\includegraphics[width=0.25\textwidth]{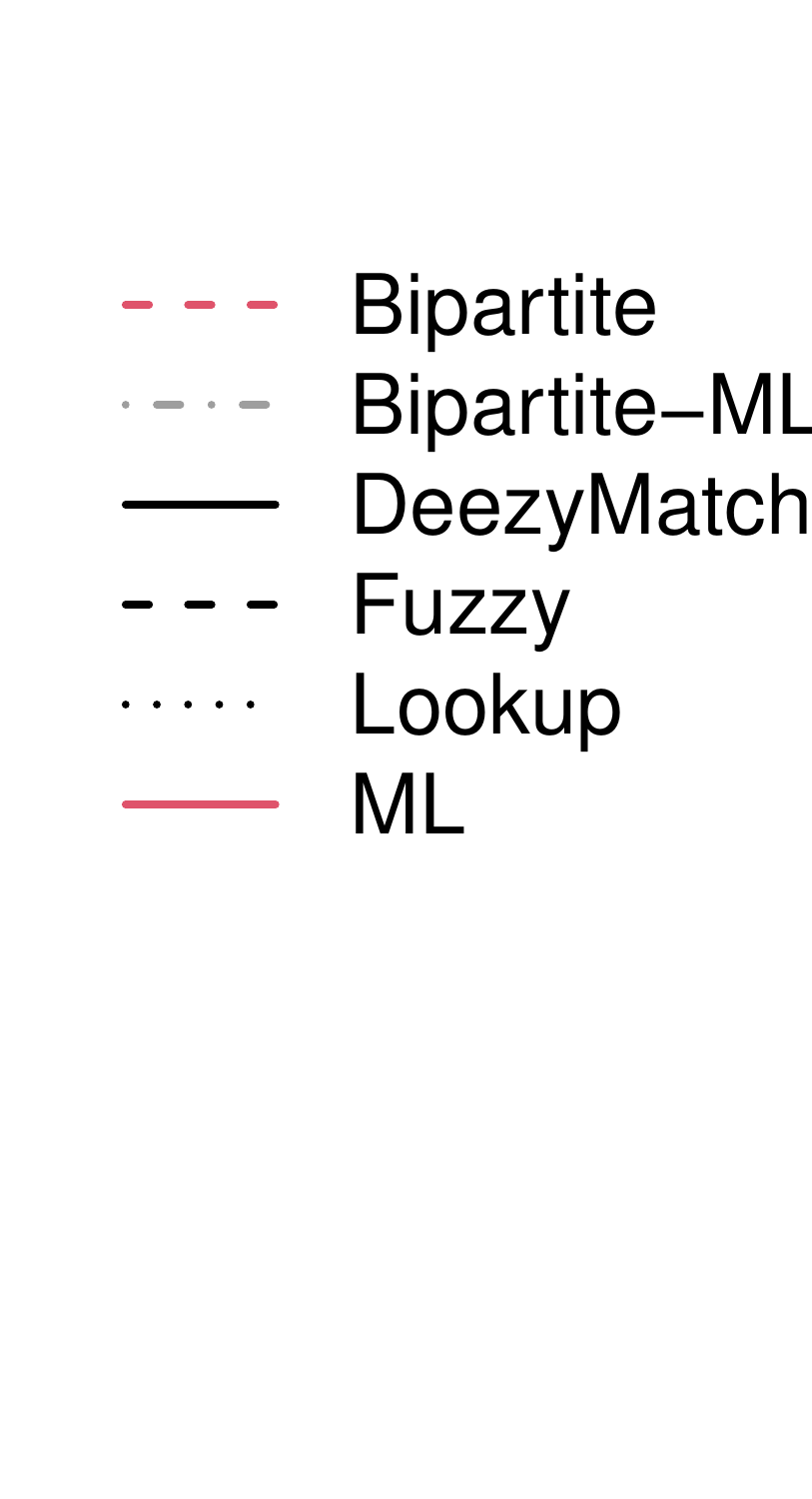}
\caption{In this YCombinator example, we see that the network-based approaches offer no relative benefit in terms of true positives when adjusting for false positives, yet the machine learning approach that uses the LinkedIn corpus performs well over fuzzy matching. Higher values along the $Y$-axis are better. See Figure \ref{fig:YCombFigAppendix} for full results with Markov approaches included. 
\label{fig:YCombFig}}
\end{center}
\end{figure}

\section{Discussion: Limits and Future of the LinkedIn Data in Improving Record Linkage}\label{s:Discussion}

We have shown how to use half a billion user-contributed records from a prominent employment networking site to help link datasets about organizations. Researchers studying organizations frequently find themselves in situations where they must link datasets based on shared names and without common covariates \citep{Crosson2020,Thieme2020,Carpenter2022,Stuckatz2022,GonzalezYou2024,Rasmussen2021ExecutiveRevolvingDoor,AbiHassan2023Ideologies}. Existing methods, notably human coding and fuzzy matching, or some combination of the two, are costly to apply and often involve ad hoc decision-making by scholars about what seems to be working well (or well enough). We have shown how the LinkedIn corpus can be used, either via machine learning or network detection, to improve organizational record linkage. These approaches are summarized in Table \ref{tab:ModelComparison}.



\begin{table}[ht]
\centering
\renewcommand\arraystretch{1.5}
\caption{Comparing different approaches to organizational record linkage.}\label{tab:ModelComparison}
\begin{tabular}[t]{
>{\raggedright}p{0.23\linewidth}
>{\raggedright}p{0.15\linewidth}
>{\raggedright}p{0.15\linewidth}
>{\raggedright\arraybackslash}p{0.15\linewidth}
>{\raggedright\arraybackslash}p{0.15\linewidth}
>{\raggedright\arraybackslash}p{0.35\linewidth}
}
\toprule\midrule 
& {\it Fuzzy String Matching} & {\it LinkedIn-Calibrated ML} & {\it LinkedIn Network Approaches} & {\it Combined ML+Network Approach}
\\ \midrule
\underline{\it Character} &  &  &  &
\\  Optimized for organizational name matching? & No  & Yes & No & Partially
\\ Text representation & Discrete & Continuous  & Discrete & Continuous
\\ Information used& Semantic & Semantic & Graph theoretic & Semantic + graph theoretic
\\ Hyper-parameters& Acceptance threshold; $q$ -gram settings& Acceptance threshold; ML model architecture & Acceptance threshold; $q$-gram settings; clustering hyperparameters & Acceptance threshold; ML model architecture; clustering hyperparameters
\vspace{0.4cm}
\\ \underline{\it Data Requirements} &  &  &  & 
\\ Requires access to saved matching model parameters? & No & Yes & No & Yes
\\ Requires access to saved alias clustering? & No & No &Yes & Yes
\\ \bottomrule
\end{tabular}
\end{table}%


These results may have implications for applied work. In our second application, the choice of record linkage method is consequential for the ultimate regressions that one runs and intends to present to other scholars. Using a unified approach, we were able to estimate a coefficient of theoretical interest within a 95\% confidence interval using ground truth. Using other methods, particularly fuzzy matching, we were unable to recover the coefficient of interest. Although the sign was correct, the magnitude was statistically and substantively different.

Typically, scholars do not have access to ground truth and, therefore, will not have a sense of how well or how badly they are doing in the aggregate. This is a potentially serious problem affecting research on organizations; however, we do not believe that this application alone should cast substantial doubt on what scholars have been doing. Typically, researchers use a mix of hand-coding and automated methods, and we expect that this kind of approach will do better than a purely automated approach (especially one relying on string distance metrics alone). 
For linkage problems that are too big for mixed workflows ($>10^5$ observations), the work here suggests it is important to test sensitivity to linkage and hyperparameter choice. We provide some examples of how that might be done.

While the integration of the LinkedIn corpus here would seem to improve organizational match performance on real data tasks, there are many avenues for future extensions in addition to those already mentioned.

First, to incorporate auxiliary information and to adjust for uncertainty about merging in post-merge analyses, probabilistic linkage models are an attractive option for record linkage tasks on individuals \citep{Enamorado2019}. In such models, a latent variable indicates whether a pair of records does or does not represent a match, inferred via Expectation Maximizing using information about the agreement level for a set of variables, such as birth date, name, residence, and, potentially, employer. Information from these LinkedIn-assisted algorithms can be readily incorporated into these algorithms for estimating match probabilities on individuals.

The methods described here might also incorporate covariate information about companies. For instance, researchers can incorporate such information in the final layer of the LinkedIn-based machine learning model and re-train that layer using a small training corpus. This process, an application of transfer learning, enables extra information to be brought to bear while also retaining the rich numerical representations obtained from the original training process performed on the massive LinkedIn dataset. Finally, the approaches here are complementary to those described in \citet{kaufman_klevs_2021}, and it would be interesting to explore possible combined performance gains. 

\section{Conclusion}\label{s:Conclusion}

Datasets that are important to scholars of organizational politics often lack common covariate data. This lack of shared information makes it difficult to apply probabilistic linkage methods and motivates the widespread use of fuzzy matching algorithms. Yet fuzzy matching is often an inadequate tool for the task at hand, while human coding is frequently costly, particularly if one wants human coders with the specialized domain knowledge necessary to generate high-quality matches. We have introduced a novel data source for improving the matching of organizational entities using half a billion open-collaborated employment records from a prominent online employment network. We show how this data can be used to match organizations that contain no common words or even characters.

We validate the approach on example tasks. We show favorable performance to the most common alternative automated method (fuzzy matching), with gains of up to 60\%. We also illustrated how increased match quality can yield improved substantive insights and better statistical precision and predictive accuracy. Our primary contribution to the research community is providing a data source that can, in ways explored here and hopefully refined in future work, improve organizational record linkage while using this unique and useful corpus. \hfill \(\square\)

\clearpage \newpage 

\printbibliography

@article{Crosson2020,
  title={Polarized Pluralism Organizational Preferences and Biases in the American Pressure System},
  author={Crosson, Jesse M and Furnas, Alexander C and Lorenz, Geoffrey M},
  journal={American Political Science Review.},
  volume={114},
  number={4},
  pages={1117–-1137},
  year={2020}
}

@article{chen2015corporate,
  title={{Corporate Lobbying and Firm Performance}},
  author={Chen, Hui and Parsley, David and Yang, Ya-Wen},
  journal={Journal of Business Finance \& Accounting},
  volume={42},
  number={3-4},
  pages={444--481},
  year={2015},
  publisher={Wiley Online Library}
}

@article{clauset2004finding,
  title={{Finding Community Structure in Very Large Networks}},
  author={Clauset, Aaron and Newman, Mark EJ and Moore, Cristopher},
  journal={Physical Review E},
  volume={70},
  number={6},
  pages={},
  year={2004},
  publisher={APS}
}

@article{mikolov2013distributed,
  title={{Distributed Representations of Words and Phrases and Their Compositionality}},
  author={Mikolov, Tomas and Sutskever, Ilya and Chen, Kai and Corrado, Greg and Dean, Jeffrey},
  journal={arXiv preprint arXiv:1310.4546},
  year={2013}
}

@article{Figlio2014a,
author = {Figlio, David and Guryan, Jonathan and Karbownik, Krzysztof and Roth, Jeffrey},
doi = {10.1257/aer.104.12.3921},
issn = {00028282},
journal = {American Economic Review},
number = {12},
pages = {4205--4230},
title = {{The Effects of Poor Neonatal Health on Children's Cognitive Development?}},
volume = {104},
year = {2014}
}

@article{Bolsen2014,
author = {Bolsen, Toby and Ferraro, Paul J. and Miranda, Juan Jose},
doi = {10.1111/ajps.12052},
issn = {15405907},
journal = {American Journal of Political Science},
number = {1},
pages = {17--30},
title = {{Are Voters More Likely to Contribute to Other Public Goods? Evidence From a Large-scale Randomized Policy Experiment}},
volume = {58},
year = {2014}
}

@article{van2008graph,
author = {{Van Dongen}, Stijn},
journal = {SIAM Journal on Matrix Analysis and Applications},
number = {1},
pages = {121--141},
title = {{Graph Clustering Via a Discrete Uncoupling Process}},
volume = {30},
year = {2008}
}

@article{rohe2011spectral,
author = {Rohe, Karl and Chatterjee, Sourav and Yu, Bin},
journal = {The Annals of Statistics},
number = {4},
pages = {1878--1915},
title = {{Spectral Clustering and the High-Dimensional Stochastic Blockmodel}},
volume = {39},
year = {2011}
}

@article{Libgober2020a,
author = {Libgober, Brian},
journal = {Quarterly Journal of Political Science},
title = {{Meetings, Comments, and the Distributive Politics of Rulemaking}},
year = {2020}
}

@article{Enamorado2019,
author = {Enamorado, Ted and Fifield, Benjamin and Imai, Kosuke},
doi = {10.1017/S0003055418000783},
isbn = {0003055418},
issn = {15375943},
journal = {American Political Science Review},
number = {2},
pages = {353--371},
title = {{Using a Probabilistic Model to Assist Merging of Large-scale Administrative Records}},
volume = {113},
year = {2019}
}

@article{Herzog2010,
author = {Herzog, Thomas H. and Scheuren, Fritz and Winkler, William E.},
doi = {10.1002/wics.108},
issn = {19395108},
journal = {Wiley Interdisciplinary Reviews: Computational Statistics},
number = {5},
pages = {535--543},
title = {{Record Linkage}},
volume = {2},
year = {2010}
}

@article{Hill2017,
author = {Hill, Seth J. and Huber, Gregory A.},
doi = {10.1007/s11109-016-9343-y},
issn = {01909320},
journal = {Political Behavior},
number = {1},
pages = {3--29},
publisher = {Springer US},
title = {{Representativeness and Motivations of the Contemporary Donorate: Results from Merged Survey and Administrative Records}},
volume = {39},
year = {2017}
}

@article{Larsen2001,
author = {Larsen, Michael D. and Rubin, Donald B.},
doi = {10.1198/016214501750332956},
issn = {1537274X},
journal = {Journal of the American Statistical Association},
number = {453},
pages = {32--41},
title = {{Iterative Automated Record Linkage Using Mixture Models}},
volume = {96},
year = {2001}
}

@article{kaufman_klevs_2021, title={Adaptive Fuzzy String Matching: How to Merge Datasets with Only One (Messy) Identifying Field}, DOI={10.1017/pan.2021.38}, journal={Political Analysis}, publisher={Cambridge University Press}, author={Kaufman, Aaron R. and Klevs, Aja}, year={2021}, pages={1–7}}

@article{lever2016classification,
  title={{Classification Evaluation: It Is Important to Understand Both What a Classification Metric Expresses and What It Hides}},
  author={Lever, Jake},
  journal={Nature Methods},
  volume={13},
  number={8},
  pages={603--605},
  year={2016},
  publisher={Nature Publishing Group}
}

@article{devarriya2020unbalanced,
  title={{Unbalanced Breast Cancer Data Classification Using Novel Fitness Functions in Genetic Programming}},
  author={Devarriya, Divyaansh and Gulati, Cairo and Mansharamani, Vidhi and Sakalle, Aditi and Bhardwaj, Arpit},
  journal={Expert Systems with Applications},
  volume={140},
  pages={112866},
  year={2020},
  publisher={Elsevier}
}

@article{eun2021aspirations,
  title={{Aspirations and Corporate Lobbying in the Product Market}},
  author={Eun, Jihyun and Lee, Seung-Hyun},
  journal={Business \& Society},
  volume={60},
  number={4},
  pages={844--875},
  year={2021},
  publisher={SAGE Publications Sage CA: Los Angeles, CA}
}

@article{ridge2017beyond,
  title={{Beyond Lobbying Expenditures: How Lobbying Breadth and Political Connectedness Affect Firm Outcomes}},
  author={Ridge, Jason W and Ingram, Amy and Hill, Aaron D.},
  journal={Academy of Management Journal},
  volume={60},
  number={3},
  pages={1138--1163},
  year={2017},
  publisher={Academy of Management Briarcliff Manor, NY}
}

@misc{OpenSecrets,
  title = {Open Secrets},
  year={2022},
  howpublished = {\url{opensecrets.org/}},
  note = {Accessed: 2022-01-01}
}

@article{agrawal2022large,
  title={{Large Language Models are Zero-shot Clinical Information Extractors}},
  author={Agrawal, Monica and Hegselmann, Stefan and Lang, Hunter and Kim, Yoon and Sontag, David},
  journal={arXiv preprint arXiv:2205.12689},
  year={2022}
}

@article{wei2022emergent,
  title={{Emergent Abilities of Large Language Models}},
  author={Wei, Jason and Tay, Yi and Bommasani, Rishi and Raffel, Colin and Zoph, Barret and Borgeaud, Sebastian and Yogatama, Dani and Bosma, Maarten and Zhou, Denny and Metzler, Donald and others},
  journal={arXiv preprint arXiv:2206.07682},
  year={2022}
}

@misc{Microsoft2016AcquireLinkedIn,
  author = {{Microsoft News Center}},
  title = {{Microsoft to Acquire LinkedIn}},
  year = {2016},
  howpublished = {\url{https://news.microsoft.com/2016/06/13/microsoft-to-acquire-linkedin/}},
  note = {}
}

@inproceedings{hosseini2020deezymatch,
  title={{DeezyMatch: A Flexible Deep Learning Approach to Fuzzy String Matching}},
  author={Hosseini, Kasra and Nanni, Federico and Ardanuy, Mariona Coll},
  booktitle={Proceedings of the 2020 Conference on Empirical Methods in Natural Language Processing: System Demonstrations},
  pages={62--69},
  year={2020}
}

@article{ruggles2018historical,
  title={{Historical Census Record Linkage}},
  author={Ruggles, Steven and Fitch, Catherine A and Roberts, Evan},
  journal={Annual Review of Sociology},
  volume={44},
  pages={19--37},
  year={2018},
  publisher={Annual Reviews}
}

@article{jiang2023mistral,
  title={Mistral 7B},
  author={Jiang, Albert Q and Sablayrolles, Alexandre and Mensch, Arthur and Bamford, Chris and Chaplot, Devendra Singh and Casas, Diego de las and Bressand, Florian and Lengyel, Gianna and Lample, Guillaume and Saulnier, Lucile and others},
  journal={arXiv preprint arXiv:2310.06825},
  year={2023}
}

@online{Goh2022,
  author  = {Steven Goh},
  title   = {{LinkDB - Exhaustive Dataset of LinkedIn People \& Company Profiles}},
  year    = {2022},
  url     = {https://nubela.co/blog/linkdb-an-exhaustive-dataset-of-linkedin-members-and-companies/},
  urldate = {2024-03-02},
  month   = sep,
  day     = {27},
  publisher = {Nubela},
  note    = {Accessed: 2024-03-02}
}

@article{rodriguez2022word,
  title={{Word Embeddings: What Works, What Doesn’t, and How to Tell the Difference for Aplied Research}},
  author={Rodriguez, Pedro L and Spirling, Arthur},
  journal={The Journal of Politics},
  volume={84},
  number={1},
  pages={101--115},
  year={2022},
  publisher={The University of Chicago Press Chicago, IL}
}

@article{GonzalezYou2024,
  author    = {Juan Pablo González and Hye Young You},
  title     = {Money and Cooperative Federalism: Evidence from EPA Civil Litigation},
  journal   = {Journal of Law, Economics, \& Organization},
  year      = {2024},
  note      = {Forthcoming}
}

@article{Thieme2020,
  author    = {Sebastian Thieme},
  title     = {Moderation or Strategy? Political Giving by Corporations and Trade Groups},
  journal   = {The Journal of Politics},
  volume    = {82},
  number    = {3},
  pages     = {1171--1175},
  year      = {2020},
  doi       = {10.1086/707619},
}

@article{Stuckatz2022,
  author    = {Jan Stuckatz},
  title     = {How the Workplace Affects Employee Political Contributions},
  journal   = {American Political Science Review},
  volume    = {116},
  number    = {1},
  pages     = {54--69},
  year      = {2022},
  doi       = {10.1017/S0003055421000836},
}

@article{Rasmussen2021ExecutiveRevolvingDoor,
  title={The Executive Revolving Door: New Dataset on the Career Moves of Former Danish Ministers and Permanent Secretaries},
  author={Rasmussen, A. Bbitex and Buhmann-Holmes, N. and Egerod, B.C.K.},
  journal={Scandinavian Political Studies},
  volume={44},
  pages={487--502},
  year={2021},
  publisher={Wiley},
  doi={10.1111/1467-9477.12214}
}

@article{AbiHassan2023Ideologies,
  title={The Ideologies of Organized Interests and Amicus Curiae Briefs: Large-Scale, Social Network Imputation of Ideal Points},
  author={Abi-Hassan, S. and Box-Steffensmeier, J.M. and Christenson, D.P. and Kaufman, A.R. and Libgober, B.},
  journal={Political Analysis},
  volume={31},
  number={3},
  pages={396--413},
  year={2023},
  publisher={Cambridge University Press},
  doi={10.1017/pan.2022.34}
}

@inproceedings{Carpenter2022,
  title={Inequality in Administrative Democracy: Large-Sample Evidence from American Financial Regulation},
  author={Daniel Carpenter and Angelo Dagonel and Devin Judge-Lord and Christopher T. Kenny and Brian Libgober and Jacob Waggoner and Steven Rashin and Susan Webb Yackee},
    organization={American Political Science Association Annual Conference},
  year={2021}
}

@article{green2023zoomerjoin,
  title={Zoomerjoin: Superlatively-Fast Fuzzy Joins},
  author={Green, Beniamino},
  journal={Journal of Open Source Software},
  volume={8},
  number={89},
  pages={5693},
  year={2023}
}



\section*{Supplementary material}

# Contents

# Appendix I: Machine Learning Model Details

We here provide details justifying the probabilistic outcome used in the machine learning approach.

## A.I.1.1 Deriving Linkage Outcome Data via LinkedIn

We first assume that the alias-to-URL pairings are relevant for understanding alias-to-organization ties. That is, we assume validity:

$$\text{Validity Assumption: } \Pr(O_i = O_j \mid A_i = a, A_j = a') = \Pr(U_i = U_j \mid A_i = a, A_j = a'), \quad (1)$$

where $O_i$, $O_j$ refer to the organizations associated with index $i$ and $j$, $A_i$, $A_j$ refer to the aliases, and $U_i$, $U_j$ refer to the URLs. Under this assumption, the overall behavior of users linking their aliases to organizational URLs is the same as if they were linking to actual organizations. Intuitively, the assumption means that the LinkedIn name-to-URL patterns give us information about the name-to-organization mapping.

Because, in linkage tasks, the organizational unit of analysis is defined by the researcher, this assumption may be satisfied in some circumstances but not in others. For example, General Motors has divisions called Buick and Cadillac, which were initially separate companies. A researcher might want these entities to match each other in some datasets. Other researchers may not want these to match, however, and for plenty of purposes, it may not make sense to make these entities connect. Therefore, whether alias-to-URL linkage is valid for a particular alias-to-organization match will depend on the way in which organizations are defined in the overall analysis.

Next, in order to use the LinkedIn corpus for learning about how different organizational names are related, we assume semantic mapping, which states that the organizational match probability is a function of the semantic content of the aliases. That is,

$$\text{Semantic Mapping Assumption: } \Pr(O_i = O_j \mid A_i, A_j) = f_p(f_n(A_i), f_n(A_j)), \quad (2)$$

where $f_n$ represents a function that numericizes each alias string, and $f_p$ represents a function that transforms the alias numeric representations into overall match probabilities. Semantic mapping, which depends on the existence of a match function, is not insignificant given the many possibilities for the researcher-defined organizational unit of analysis. It also could be questionable for cross-language link tasks. Nevertheless, assuming that this function exists, we can attempt to approximate it using modern machine-learning algorithms in combination with the trillions of LinkedIn organizational alias pairs.

In clarifying the limitations of the LinkedIn data for improving linkage, we note that both the validity and semantic mapping assumptions would be vulnerable to errors in the LinkedIn data. Regardless of how researchers define their organizational unit of analysis, inaccuracies due to the selection of incorrect URLs by users would limit our ability to capture the link probability using the alias and URL information. It could introduce systematic bias as well. However, as a professional networking site, users have an incentive to maintain accurate information about themselves on the page, hopefully limiting the degree of systematic bias.

Finally, to generate ground truth data for the prediction model training, we assume independence of $i$ and $j$, which allows us to calculate $\Pr(O_i = O_j \mid A_i, A_j)$. This independence assumption would be violated if, for example, users could coordinate so that one use of an alias linking to a given URL gave additional information about another use. Given the wide geographic reach of the LinkedIn network, this sort of coordination is presumably rare, yet is difficult to evaluate for all pairs of

indices. In any case, the assumption is important because, using it, we can calculate the overall organizational link probability using the LinkedIn data:

$$\Pr(O_i = O_j \mid A_i = a, A_j = a') = \Pr(U_i = U_j \mid A_i = a, A_j = a') \text{ (by \textit{Validity})} \tag{3}$$

$$= \sum_{u \in \mathcal{U}} \Pr(U_i = u, U_j = u \mid A_i = a, A_j = a') \tag{4}$$

$$= \sum_{u \in \mathcal{U}} \Pr(U_i = u \mid A_i = a) \Pr(U_j = u \mid A_j = a') \text{ (by \textit{Independence})} \tag{5}$$

The two terms in the final expression can be calculated using the LinkedIn network with empirical frequencies (e.g., the fraction of times alias $a$ and $a'$ link to $u$).

#### A.I.1.1.1 Modeling Approaches

There are many estimation models consistent with the core assumptions we describe above. Indeed, as machine-learning methods continue to evolve, the future will no doubt yield improvements in any machine-learning approach to modeling these match probabilities, as the rapid progress in large language models has made apparent (Jiang et al., 2023; Wei et al., 2022).

We here illustrate use of the LinkedIn data using a modeling strategy that explicitly optimizes match probabilities. For computational reasons which will soon become clear, we assume the following functional form for the match probabilities:

$$\textit{Distance-to-Probability Mapping: } \log\left(\frac{\Pr(O_i = O_j \mid A_i = a, A_j = a')}{1 - \Pr(O_i = O_j \mid A_i = a, A_j = a')}\right) \propto -||f_n(a) - f_n(a')||_2. \tag{6}$$

Here, the match probability is a function of the Euclidean distance between the numerical representations for aliases $a$ and $a'$. This captures the intuition that intuitively close names like "apple" and "apple inc." are more likely to be matched than distant names like "apple" and "jp morgan".

Besides this intuition, this distance-to-probability mapping also has important computational benefits: if we seek to calculate the match probability between two datasets of 100 observations each, we do not have to apply a computationally expensive non-linear $f_p$ function to all 5,050 possible pairs. Instead, we only need to apply the numericization function 200 times—once for each alias in each dataset—after which we take the Euclidean distance between each pair of numerical representations in a step that can be readily parallelized and optimized for speed (Assis Zampirolli and Filipe, 2017; Hansen, 2007).

### A.I.1.2 Model Implementation Details

The character vectors and neural weights are model parameters, which are jointly updated using stochastic gradient descent to minimize prediction error on our training data computed using the LinkedIn corpus. Specifically, we minimize the KL divergence between the true match probability, $\Pr(O_i = O_j | A_i = a, A_j = a')$, and the estimated probability, where the KL divergence between two distributions, $P$ and $Q$, denoted by $D_{\text{KL}}(P||Q)$, is a measure of how much information $Q$ contains that $P$ does not and is 0 when the two distributions contain the same information. That is, we minimize:

$$\text{minimize} \sum_{a,a' \in \mathcal{A}} D_{\text{KL}}\left(\widehat{\Pr}(O_i = O_j | A_i = a, A_j = a') \;||\; \Pr(O_i = O_j | A_i = a, A_j = a')\right), \tag{7}$$

where $\mathcal{A}$ denotes the set of all aliases. In this way, we learn from data how the aliases should be numerically represented—learning along the way the representation of characters, words, and how these things combine together in a name—with the overall goal of predicting the organizational link probability between each alias pair. We apply the approach here jointly to the entire LinkedIn corpus so that name representations in many languages are learned together.

We now discuss in more detail the machine learning model used.

### A.I.1.3 Step 1. Obtaining Organizational Name Representations

We first outline the machine learning modeling strategy for each organizational name.

*Input*: An organizational name (e.g., "j.p. morgan chase")

1. Break each name into words.

    - For each word of length $l_w$, query the vector representation for each character to form a $l_w \times D$-dimensional representation for that word.
    - Apply a Transformer model to each word to obtain a representation for each word.

2. For each alias of $l_a$ words, query the word vectors obtained from the previous step to form a $l_a \times D$-dimensional representation for the alias.

    - Apply a Transformer model to each alias to obtain a representation for each sequence of words.

3. Select a [CLS]-token representing the alias-level representation. Normalize and project to the final output dimensionality.

*Output*: $D$-dimensional representation of the organizational name

### A.I.1.4 Step 2. Transforming Name Representations into Match Probabilities

With the name representations in hand, we now discuss in greater detail how we form the final match probabilities.

*Input*: Numerical representations/embeddings for two organizational names

1. Take the normalized Euclidean distance between the two embeddings, $e_i$, $e_j$:

$$d_{ij} = \frac{||e_i - e_j||_2}{\sqrt{\dim(e_i)}}.$$

2. Form the match probability from the distance using unidimensional logistic regression:

$$\Pr(O_i = O_j | A_i = a, A_j = a') = \frac{1}{1 + \exp(-[\beta_0 + \beta_1 d_{ij}])}.$$

*Output*: A match probability indicating the probability that two names refer to the same organization

The normalization layers are included to avoid exploding gradients (e.g., large values are centered and scaled) and to make the learning more robust to initialization (Santurkar et al., 2018). The parameters used in steps 1 and 2 are updated jointly using a modified form of stochastic gradient descent. (Future progress can likely be made here, given the development of large language models involving billions of parameters.) We next give more details about the sampling procedure involved in the model training.

### A.I.1.5 Sampling Design for Model Training

Recall that we seek to optimize the following:

$$\text{minimize} \sum_{a,a' \in \mathcal{A}} D_{\text{KL}} \left( \widehat{\Pr}(O_i = O_j | A_i = a, A_j = a') \,||\, \Pr(O_i = O_j | A_i = a, A_j = a') \right).$$

One challenge in performing this optimization in practice is that most $a'$ are very distinct semantically from $a$, so the learning process is slowed by the vast number of "easy" predictions (e.g., where $\Pr(O_i = O_j | A_i, A_j)$ is clearly 0). To make matters worse, most match probabilities are 0, a kind of imbalance that can impair final model performance (Kuhn and Johnson, 2013)

We adopt two approaches to address these challenges. First, we implement a balanced sampling scheme: in the optimization, we ensure half of all training points have $\Pr(O_i = O_j | A_i = a, A_j = a') = 0$ and half have $\Pr(O_i = O_j | A_i = a, A_j = a') > 0$.

Second, we select $a, a'$ pairs that will be maximally informative so that the model is able to more quickly learn from semantically similar non-matches (e.g., "The University of Chicago" from "The University of Colorado") or semantically distinct matches (e.g., "Massachusetts Institute of Technology" and "MIT"). In our adversarial sampling approach (which is somewhat similar to (Kim et al., 2020)), we select, for non-matches in our training batch, alias pairs that have $\Pr(O_i = O_j | A_i = a, A_j = a') = 0$ but have the largest $\widehat{\Pr}(O_i = O_j | A_i = a, A_j = a')$. Likewise, for matches, we select pairs that have the lowest predicted probability. More formally, we find the negative $a$ and $a'$ pairs by solving:

$$\text{argmax}_{a,a' \text{ s.t. } \Pr(O_i=O_j|A_i=a,A_j=a')=0} D_{\text{KL}} \left( \Pr(O_i = O_j | A_i = a, A_j = a') \,||\, \widehat{\Pr}(O_i = O_j | A_i = a, A_j = a') \right),$$

with a similar approach applied to positive pairs.

We achieve this in practice by, at the current state of the model, finding the closest alias to $a'$ in the alias vector space for which $\Pr(O_i = O_j | A_i = a, A_j = a')$ is 0. A similar approach is used to obtain high-information positive matches. Intuitively, this process magnifies the importance of similar aliases that refer to different organizations in the learning process so as to learn how to model the semantic mapping of aliases into match probabilities.

We found this adaptive sampling scheme to be important in learning to quickly distinguish similar aliases that refer to different organizations, although only use this adversarial approach to obtain half of each training batch to mitigate against potential problems with this approach (such as occurs if the model is fit on only high surprise mistake alias pairs that occur when users incorrectly link to the same URL).

# Appendix II: Additional Community Detection Details

The first approach towards organizational name community detection uses Markov clustering (Van Dongen, 2008). The Markov clustering algorithm operates on weighted adjacency matrices that have a probabilistic interpretation.

We denote the organization names in the source data as aliases $a$. These aliases constitute nodes on our graph. Many aliases appear simultaneously with profile URLs in our source data. One possible approach to defining links between aliases is to assume that two aliases are connected whenever they both link to the same profile URL. However, this approach would place as strong a connection on aliases that rarely link to the same profile URL as those that are frequently connected. For example, the URL `linkedin.com/company/university-of-michigan` has been associated with an alias 75,462 distinct times in our data. The vast majority of these connections occur via the alias "University of Michigan", but a link with the alias "Michigan State University" does in fact occur a handful of times. In defining links on the network, it will help to upweight the former but downweight the latter in a data-driven way.

While there are many possible ways to ensure more frequently made connections are weighted higher than connections that are rarer, we adopt an approach inspired by naive Bayesian classification methods. We define the edge weight, $w_{aa'}$, between node $a$ and $a'$ using Equation 8:

$$w_{aa'} = \sum_{u \in \mathcal{U}} \frac{\text{\# of times } a \text{ co-occurs with } u}{\text{\# of times } u \text{ co-occurs with any alias}} \times \frac{\text{\# of times } u \text{ co-occurs with alias } a'}{\text{\# of times } a' \text{ occurs with any profile URL}} \tag{8}$$

These edge weights then make up the full adjacency matrix, $\mathbf{W}$, which captures the interrelationship between all pairs of aliases.

The connection of this expression to naive Bayesian classifiers requires some elaboration. As a pure formalism, one can write the law of total probability as

$$\Pr(A_i = a \mid A_j = a') = \sum_{u \in \mathcal{U}} \Pr(A_i = a \mid U = u, A_j = a') \Pr(U = u \mid A_j = a') \tag{9}$$

Taking this equation as given, suppose further one were to "naively" assume conditional independence of $A_i$ and $A_j$ given $U$ (Rish et al., 2001), or more formally $\Pr(A_i = a \mid U = u, A_j = a') = \Pr(A_i = a \mid U = u)$, which is similar to the independence assumption of Section A.I.1.1.1.[1] Equation 8 is then what would result if one also replaced the true probabilities with sample proportions.

There are several benefits to using this formula, especially over the more obvious "shared link" approach. Besides using much more of the data, the naive Bayesian calculation reweights edges in ways that are proportionate to the actual prevalence of relationships found in the data. One corollary is that the edge weights are asymmetric, and the specificity with which an alias and link are shared matters a great deal. For example, ICPSR is a rare alias, but when it does occur, it is very often entered with `linkedin.com/company/university-of-michigan`, which in turn usually points to the "University of Michigan" alias. Therefore, the weight $w_{\text{"University of Michigan","ICPSR"}}$ will be close to 1 even if $w_{\text{"ICPSR","University of Michigan"}}$ is very small. By contrast, since "Michigan State University" and "University of Michigan" are both relatively common aliases, one spurious link

---

[1] By asserting the equation as a formalism, we do not dwell on questions about what exactly the probability of $A_i$ given $A_j$ means in this context, which might distract us from the core task of showing how community detection methods on networks can be useful for record linkage problems.

between them will not result in large weight in either direction.

Having built an adjacency matrix, **W**, the probabilistic network of alias-to-alias links, we next apply the Markov clustering algorithm to it to find communities of aliases that tend to link to the same URLs. Because our focus is on the LinkedIn data contribution, we leave details to Section A.II.1.1.

Figure A.II.1 illustrates the clustering process on a subset of our data. Darker shades reflect heavier weights at initialization. Some links are much stronger than others. In the initial weighting, two cliques reflect a set of names associated with "JP Morgan Chase." Another reflects names associated with "Bank of America." However, these initial links are dense, making it difficult to distinguish one cluster of aliases from another. As the algorithm iterates, some links weaken and disappear while others strengthen. Eventually, each node links to exactly one other node. Notably, the final cliques contain lexicographically dissimilar nodes that do indeed belong in the same cluster. For example, the "Chase" clique contains "Wamu", "Paymenttech", and "摩根大通" which are all Chase affiliates. The "Bank of America" clique includes "Countrywide Financial" and "MBNA," both under the Bank of America umbrella. In this way, the semantic mapping assumption from Section A.I.1.1.1 has been weakened; graph-theoretic information has been exploited to assist organizational matching.

### A.II.1.1 Markov Clustering Algorithmic Details

Briefly, this clustering algorithm proceeds in two steps—expansion and inflation—that are repeated until convergence. In the expansion step, the network's adjacency matrix is multiplied by itself $k$ times. This matrix multiplication simulates the diffusion of a Markov process on the nodes (i.e. "traveling" $k$ steps on the network, with probabilities of where to go defined by **W**). In the inflation step, entries of the resulting matrix are raised to some power $p$, and the matrix is renormalized into a valid Markov transition matrix. Since small probabilities shrink faster under exponentiation than large probabilities, the inflation step causes higher probability states to stand out (i.e. likely places are made even more likely). After alternating expansion and inflation, the output converges: row $i$ will have only one non-zero valued entry in column $j$, which defines the representative node in the community. All rows that have a one in column $j$ are a part of the same community of alias linking to (hopefully) the same organizational entity.

This clustering process involves hyperparameters, in particular, the number of matrix multiplications to do in the expansion step ($k$) and the power of exponentiation ($p$) in the inflation step. For both steps, $k$ and $p$ are frequently set to 2. The total number of clusters is not explicitly specified but instead controlled indirectly through $p$ and $k$. We adopt the common choice of $p = k = 2$.[2]

### A.II.1.2 Greedy Bipartite Clustering Details

Following Clauset et al. (2004), our modularity score for bipartite community detection via greedy clustering is

$$\text{Modularity Score} = \frac{1}{2m} \sum_{a,u} \left[ B_{au} - \frac{k_a k_u}{2m} \right] \mathbb{I}\{c_a = c_u\}, \tag{10}$$

where $m$ denotes the total number of edges in the graph, $B_{au}$ denotes the $au$-th entry in the weighted bipartite network (i.e., the number of ties between alias $a$ and URL $u$), and $k_a$ denotes the degree of node $a$ (i.e. $k_a = \sum_{a \in \mathcal{A}} B_{au}$). The indicator variable, $\mathbb{I}\{\cdot\}$, is 1 when the community for $a$ equals the community for $u$ (these communities are denoted as $c_a$ and $c_u$).

---

[2]Alternative choices yielded substantially similar results in the applications that follow.

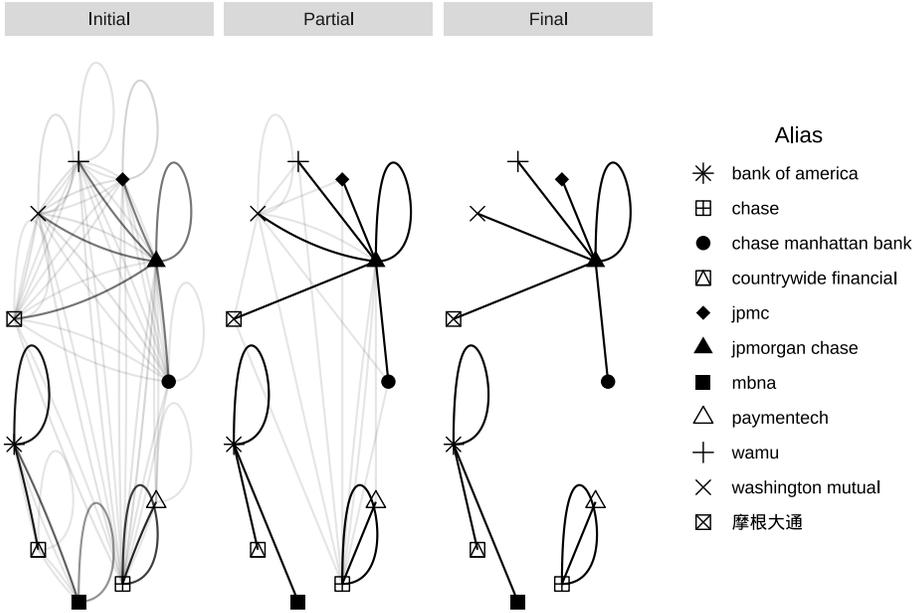

**Figure A.II.1:** *Illustration of Markov clustering.*

The intuition in Equation 10 is that the modularity score is maximized when the number of ties between $a$ and $u$ is large (i.e., $B_{au}$ is big) and $a, u$ are placed in the same cluster (so that $c_a = c_u$). We obtain community estimates based on the greedy optimization of Equation 10 (see Clauset et al. (ibid.) for details).

# Appendix III: Additional Fuzzy String Distance Details

## A.III.1.1 Fuzzy String Matching Details

Fuzzy string distances in our baseline linkage method are calculated as follows. Let $\tilde{a}$ and $\tilde{a}'$ denote the decomposition of organizational aliases $a$ and $a'$ into all their $q$-character combinations (known as $q$-grams). For example, if $q = 2$, "bank" would be decomposed into the set $\{$"ba","an","nk"$\}$. The Jaccard measure is then defined to be

$$d(a, a') = 1 - \frac{|\tilde{a} \cap \tilde{a}'|}{|\tilde{a} \cup \tilde{a}'|}.$$

If all $q$-grams co-occur within $a$ and $a'$, this measure returns 0. If none co-occur, the measure returns 1. If exactly half co-occur, the measure returns 0.5. Following (Navarro and Salmela, 2009), we set $q = 2$.

# Appendix IV: A Task Matching English and Non-English Company Names

In this supplementary evaluation task, we examine the performance of our approach in a particularly challenging case where we seek to match organizational entities using their names written in English and in Mandarin Chinese. This matching task is especially challenging because English and Mandarin Chinese are based on two entirely different kinds of writing systems (i.e., alphabetic and logographic).

The data for this task is from FluentU, an organization focusing on language learning (*FluentU* 2022). The organization has provided a directory of 76 companies with their English names paired with their Chinese names. Table A.IV.1 displays some of the companies found in this dataset, which is composed primarily of large multinational corporations.

| English Name | Chinese Name |
|---|---|
| amazon | 亚马逊 |
| coca cola | 可口可乐 |
| jp morgan | 摩根 |
| marlboro | 万宝路 |

***Table A.IV.1:*** *A sample of the organizational entities in the cross-language dataset. We attempt to match the pool of English names to their associated names in Mandarin Chinese.*

We compare linkage performance using our community detection algorithms based on the LinkedIn network and using our character-based machine learning and the baseline fuzzy matching approach. The fuzzy matching approach gives *no* matches (except when the fuzzy distance threshold is set to 1 so that all English names are matched with all Chinese names). This occurs because none of the English names share any Latin character combinations with any of the Chinese names.

The machine learning linkage approach based on the LinkedIn network fares somewhat better—yielding a true positive rate of 0.2 when the matched dataset size approaches 1,000. In contrast to fuzzy matching, it also achieves a non-zero $F_2$ score. However, the community detection approaches using the LinkedIn network fare better, successfully picking out 33 of the 76 matches with only five false positives. The left panel of Figure A.IV.1 shows how the Markov clustering approach somewhat outperforms the bipartite clustering, although both approaches achieve similar maximum true positive rates and $F_2$ scores just above 0.4. In this way, the network information present in the LinkedIn corpus allows us to find almost half of the true matches, even though the names to be matched are written in entirely different language systems.

Still, it is once again instructive to examine names for which the community detection approaches succeed and fail in this cross-language matching task. The approach succeeds in cases where an organization has employees who use both English and non-English organizational names while linking to the same company URL. For example, we successfully match the "Coca-cola" to "可口可乐" match, as Coca-cola has at least six Chinese offices, and employees in these offices often use Coca-Cola's Chinese name in their LinkedIn URL. However, we do not find the "lamborghini" to "兰博基尼" match. Lamborghini does operate offices in China, but its employees use the name "Lamborghini" in their LinkedIn profiles. In this way, the community-detection algorithms based

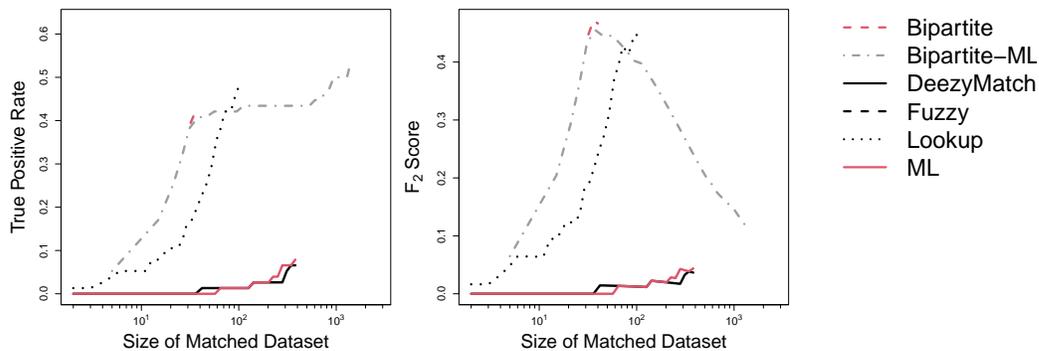

***Figure A.IV.1:*** *In the left panel, we see that the LinkedIn-based community detection algorithms find a significant fraction of the true positive matches between the English and non-English aliases. In the right panel, we see that these true positive matches are found without introducing numerous false positives (so that the $F_2$ scores for the community detection algorithms approach 0.5). The machine learning approach yields some true positive matches; fuzzy matching yields none.*

on the LinkedIn network are, for cross-language merge tasks, best suited for the matching of organizational entities that have many employees across linguistic contexts and where non-English names are commonly used on LinkedIn.

# Appendix V: A LinkOrgs Package Tutorial

The package can be downloaded via `GitHub`:

```
# Download package via github
devtools::install_github("cjerzak/LinkOrgs-software/LinkOrgs")

# load it into your R session
library(LinkOrgs)
```

There are several major functions in the package. The first is `FastFuzzyMatch`, which performs traditional string distance matching using all available CPU cores. This offers a substantial time speedup compared to the sequential application of a fuzzy distance calculator. The second function, `LinkOrgs`, is the main estimation function used in this paper to calculate the match probabilities. We also have a function, `AssessMatchPerformance`, that computes performance metrics of interest. To get help with any of these functions, one can run the following:

```
# To see package documentation, enter
# ?LinkOrgs::LinkOrgs
# ?LinkOrgs::AssessMatchPerformance
# ?LinkOrgs::FastFuzzyMatch
```

Here is the syntax for how to use the package using synthetic data:

```
# Create synthetic data to try everything out
x_orgnames <- c("apple","oracle","enron inc.","mcdonalds corporation")
y_orgnames <- c("apple corp","oracle inc","enron","mcdonalds co")
x <- data.frame("orgnames_x"=x_orgnames)
y <- data.frame("orgnames_y"=y_orgnames)

# Perform a simple merge with the package
linkedOrgs <- LinkOrgs(x = x,
                       y = y,
                       by.x = "orgnames_x",
                       by.y = "orgnames_y",
                       algorithm = "bipartite")

# Print results
print( linkedOrgs )
```

An up-to-date tutorial for the `LinkOrgs` package is available at

GitHub.com/cjerzak/LinkOrgs-software/blob/master/README.md

# Appendix VI: Algorithm Compute Times by Task

## A.VI.1.1 Matching Execution Times by Algorithm by Task in Minutes

**Table A.VI.1:** *Run time on the meetings data analysis.*

| Algorithm    | Run Time (mins) |
|--------------|-----------------|
| Bipartite    | 13.12           |
| Bipartite-ML | 251.38          |
| DeezyMatch   | 0.24            |
| Fuzzy        | 0.27            |
| Lookup       | 1.35            |
| Markov       | 8.61            |
| Markov-ML    | 113.00          |
| ML           | 1.63            |

**Table A.VI.2:** *Run time on the company lobbying data analysis.*

| Algorithm    | Run Time (mins) |
|--------------|-----------------|
| Bipartite    | 13.40           |
| Bipartite-ML | 318.44          |
| DeezyMatch   | 0.32            |
| Fuzzy        | 0.33            |
| Lookup       | 1.36            |
| Markov       | 9.47            |
| Markov-ML    | 142.50          |
| ML           | 2.09            |

**Table A.VI.3:** *Run time on the cross-language merge task.*

| Algorithm    | Run Time (mins) |
|--------------|-----------------|
| Bipartite    | 0.70            |
| Bipartite-ML | 6.17            |
| DeezyMatch   | 0.15            |
| Fuzzy        | 0.02            |
| Lookup       | 1.11            |
| Markov       | 0.29            |
| Markov-ML    | 3.52            |
| ML           | 1.39            |

**Table A.VI.4:** *Run time on the Y Combinator task.*

| Algorithm | Run Time (mins) |
|:---:|:---:|
| Bipartite | 7.63 |
| Bipartite-ML | 142.84 |
| DeezyMatch | 0.33 |
| Fuzzy | 0.15 |
| Lookup | 1.26 |
| Markov | 4.41 |
| Markov-ML | 66.16 |
| ML | 1.22 |

# Appendix VII: Additional Empirical Results

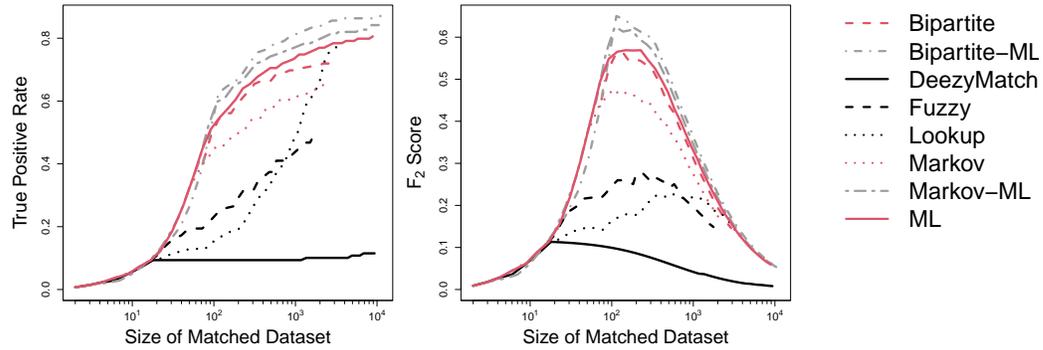

***Figure A.VII.1:*** *We find that dataset linkage using any one of the approaches using the LinkedIn network obtains favorable performance relative to fuzzy string matching both when examining only the raw percentage of correct matches obtained (left panel) and when adjusting for the rate of false positives and false negatives in the $F_2$ score (right panel). In both figures, higher values along the Y-axis are better. The "Bipartite" and "Markov" refer to two network-based approaches to linkage. "ML" refers to the machine learning approach introduced above. "Fuzzy", "DeezyMatch", and "Lookup" refer to the string distance, machine learning, and network method baseline methods. "Markov-ML" and "Bipartite-ML" refer to the ensemble approaches.*

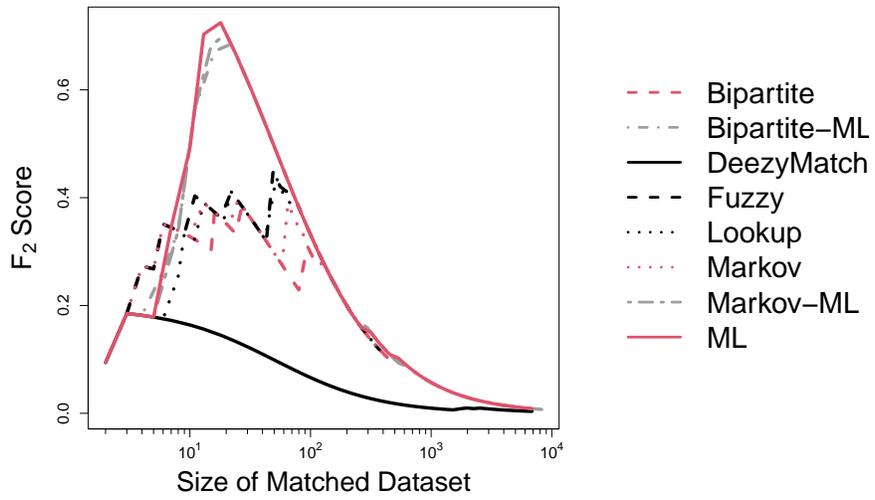

***Figure A.VII.2:*** *In this YCombinator example, we see that the network-based approaches offer no relative benefit in terms of true positives when adjusting for false positives, yet the machine-learning-assisted approaches using the LinkedIn corpus perform well over fuzzy matching. Higher values along the Y-axis are better.*

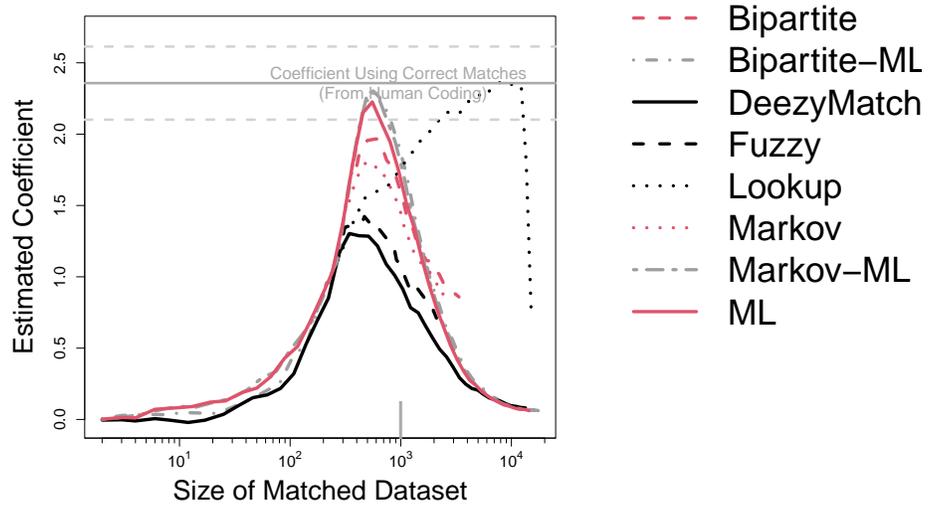

***Figure A.VII.3:*** *The coefficient on log(Assets) for predicting log(1+Expenditures) using the ground truth data is about 2.5 (bold gray line, 95% confidence interval displayed using dotted gray lines). At its best point, fuzzy matching underestimates this quantity by about half. The LinkedIn-based matching algorithms better recover the coefficient.*

17

# Appendix VIII: Data Availability Statement

The LinkedIn data corpus is available in a Harvard Dataverse:

> `doi.org/10.7910/DVN/EHRQQL`

as well as a Hugging Face repository:

> `HuggingFace.co/datasets/cjerzak/LinkOrgs`.

All methods are accessible in an open-source codebase available at

> `GitHub.com/cjerzak/LinkOrgs-software`.

# References

Assis Zampirolli, Francisco de and Leonardo Filipe (2017): "A Fast CUDA-based Implementation for the Euclidean Distance Transform". In: *2017 International Conference on High Performance Computing & Simulation (HPCS)*. IEEE, pp. 815–818.

Clauset, Aaron et al. (2004): "Finding Community Structure in Very Large Networks". In: *Physical Review E*, no. 6, vol. 70.

*FluentU* (2022). https://www.fluentu.com/blog/chinese/business-chinese-vocabulary-list-foreign-companies-in-chinese/. Accessed: 2022-01-01.

Hansen, Ben B (2007): "Optmatch: Flexible, Optimal Matching for Observational Studies". In: *New Functions for Multivariate Analysis*, no. 2, vol. 7, pp. 18–24.

Jiang, Albert Q et al. (2023): "Mistral 7B". In: *arXiv preprint arXiv:2310.06825*.

Kim, Jaekyeom et al. (2020): "Model-agnostic Boundary-adversarial Sampling for Test-time Generalization in Few-shot Learning". In: *European Conference on Computer Vision*. Springer, pp. 599–617.

Kuhn, Max and Kjell Johnson (2013): "Remedies for Severe Class Imbalance". In: *Applied Predictive Modeling*. Springer, pp. 419–443.

Navarro, Gonzalo and Leena Salmela (2009): "Indexing Variable Length Substrings for Exact and Approximate Matching". In: *International Symposium on String Processing and Information Retrieval*, pp. 214–221.

Rish, Irina et al. (2001): "An Empirical Study of the Naive Bayes classifier". In: *IJCAI 2001 Workshop on Empirical Methods in Artificial Intelligence*. Vol. 3. 22, pp. 41–46.

Santurkar, Shibani et al. (2018): "How Does Batch Normalization Help Optimization?" In: *Advances in Neural Information Processing Systems*, vol. 31.

Van Dongen, Stijn (2008): "Graph Clustering Via a Discrete Uncoupling Process". In: *SIAM Journal on Matrix Analysis and Applications*, no. 1, vol. 30, pp. 121–141.

Wei, Jason et al. (2022): "Emergent Abilities of Large Language Models". In: *arXiv preprint arXiv:2206.07682*.



\end{document}